\documentclass[12pt]{article}
\usepackage{amsmath,amssymb,amsthm,amsfonts,bm,mathrsfs,float,algorithm,algorithmic,caption,subcaption,setspace,booktabs}
\usepackage{avant,array,geometry,fontenc,inputenc,natbib,multirow,enumerate,enumitem,rotating,booktabs,color,latexsym,%times,
stmaryrd}
\usepackage{xr}
\usepackage{multirow}
\usepackage[colorlinks,linkcolor=black,citecolor=blue,urlcolor=blue]{hyperref}
\usepackage{soul}
\usepackage{tikz} % for tensor figures
\usetikzlibrary{shapes,snakes}

\makeatletter
\@ifundefined{date}{}{\date{}}

\newcolumntype{L}[1]{>{\raggedright\let\newline\\\arraybackslash\hspace{0pt}}m{#1}}
\newcolumntype{C}[1]{>{\centering\let\newline\\\arraybackslash\hspace{0pt}}m{#1}}
\newcolumntype{R}[1]{>{\raggedleft\let\newline\\\arraybackslash\hspace{0pt}}m{#1}}

\floatstyle{ruled}
\newfloat{algorithm}{tbp}{loa}
\providecommand{\algorithmname}{Algorithm}
\floatname{algorithm}{\protect\algorithmname}
\@ifundefined{definecolor}{color}{}

\newtheorem{remark}{Remark}

\newtheorem{lemma}{Lemma}
\newtheorem{corol}{Corollary}
\newtheorem{proposition}{Proposition}
\newtheorem{theorem}{Theorem}
\newtheorem{definition}{Definition}

\newtheorem*{McD*}{McDiarmid's Inequality}

\DeclareMathOperator*{\argmax}{argmax}
\DeclareMathOperator*{\argmin}{argmin}

\global\long\def\mba{\mathbf{a}}
\global\long\def\mbA{\mathbf{A}}

\global\long\def\mbp{\mathbf{p}}
 
 \global\long\def\mbB{\mathbf{B}}
  \global\long\def\mbb{\mathbf{b}}

 \global\long\def\mbx{\mathbf{x}}
 
 \global\long\def\mbu{\mathbf{u}}

 \global\long\def\mbX{\mathbf{X}}
 \global\long\def\mbY{\mathbf{Y}}

 \global\long\def\mbP{\mathbf{P}}

 \global\long\def\mbbR{\mathbb{R}}

 \global\long\def\boltheta{\boldsymbol{\theta}}
 \global\long\def\boleta{\boldsymbol{\eta}}

 \global\long\def\bolbeta{\boldsymbol{\beta}}
 
 \global\long\def\bolgamma{\boldsymbol{\gamma}}

 \global\long\def\bolSigma{\boldsymbol{\Sigma}}

\global\long\def\vecc{\mathrm{vec}}
 \global\long\def\Prob{\mathrm{Pr}}
 \global\long\def\E{\mathrm{E}}

  \global\long\def\Ptensor{\mathcal{P}_{c_1,\dots, c_M}}

\usepackage{fullpage}

\begin{document}

\title{\Large{\textbf{Conditional probability tensor decompositions\\ for multivariate categorical response regression}}}
\author{
Aaron J.~Molstad$^{\dag *}$ and Xin Zhang$^\ddag$\thanks{The authors contributed equally to this work and are listed in alphabetical order. Corresponding author: Aaron J.~Molstad (amolstad@umn.edu). Research for this paper was supported in part by grants DMS-2053697 (XZ), DMS-2113590 (XZ) and DMS-2113589 (AJM)  from the U.S.~National Science Foundation.}\\
$^\dag$University of Minnesota and $^\ddag$Florida State University
\\
}

\date{}
\maketitle

\begin{abstract}
In many modern regression applications, the response consists of multiple categorical random variables whose probability mass is a function of a common set of predictors. In this article, we propose a new method for modeling such a probability mass function in settings where the number of response variables, the number of categories per response, and the dimension of the predictor are large. Our method relies on a functional probability tensor decomposition: a decomposition of a tensor-valued function such that its range is a restricted set of low-rank probability tensors. This decomposition is motivated by the connection between the conditional independence of responses, or lack thereof, and their probability tensor rank. We show that the model implied by such a low-rank functional probability tensor decomposition can be interpreted in terms of a mixture of regressions and can thus be fit using maximum likelihood. We derive an efficient and scalable penalized expectation maximization algorithm to fit this model and examine its statistical properties. We demonstrate the encouraging performance of our method through both simulation studies and an application to modeling the functional classes of genes. \smallskip
\end{abstract}
%\newpage
\onehalfspacing

%\doublespacing
\vspace{-20pt}
\section{Introduction}
We consider the problem of modeling the conditional distribution of multiple categorical response variables as a function of a $p$-dimensional vector of predictors, i.e., a multivariate categorical response regression analysis.
%When the responses are all normal, their joint (conditional) distribution can be modeled using a likelihood-based approach. For example, 
%, the standard approach is to assume a linear conditional mean function $\E(\mbY\mid\mbX)$ and a constant conditional covariance $\Cov(\mbY\mid\mbX)$, both of which can be estimated straightforwardly via maximum likelihood. When the responses are all categorical, this and related techniques are not appropriate. 
% Because of the discrete nature of the response, special care must be taken to model the full joint (conditional) probability mass function of the responses in a flexible, interpretable, and scalable fashion.  
%When the goal is to model the probability mass of these variables as  is assumed to be a function of a common set of predictors. In particular, special care must be taken to model the full joint (conditional) probability mass function of the responses in a flexible, interpretable, and scalable fashion. In this article, we consider the multivariate response regression problem when all responses are categorical, i.e., \emph{multivariate categorical response regression}.  
To make matters concrete, for an integer $M \geq 2$, let $\mbY = (Y_1, \dots, Y_M)$ be the multivariate categorical response, where each component $Y_{m}$ has $c_m \geq 2$ many categories with numerically coded support $Y_{m} \in[c_m] = \{1,\dots,c_{m}\}$ for all $m\in[M] = \{1,\dots,M\}$. The regression models and methods developed in this article allow the predictor $\mbX \in \mathbb{R}^p$ to be either continuous or discrete (or mixed), and allow the predictors $\mbx \in\mbbR^p$ to be either random or fixed. The regression problem is essentially the study of the conditional distribution $\mbY\mid\mbX$ whose joint probability mass function consists of 
\begin{equation}\label{ProbTensor}
P_{j_{1}\dots j_{M}}(\mbx) = \Prob(Y_{1}=j_{1},\dots,Y_{M}=j_{M}\mid\mbX=\mbx)\geq0, \quad j_{m}\in[c_{m}],~~ m\in[M].
\end{equation}
From \eqref{ProbTensor}, we can define the $M${th order tensor} $\mbP(\mbx)\in\mbbR^{c_{1}\times\cdots\times c_{M}}$
whose $(j_{1},\dots,j_{M})$th element is $P_{j_{1}\dots j_{M}}(\mbx)$.
This conditional probability tensor  $\mbP(\mbx)$ fully characterizes the conditional distribution of $\mbY\mid\mbX = \mbx$ and is thus the quantity of interest in our study.

	To fit \eqref{ProbTensor}, there are numerous approaches one could consider. At one extreme, each response could be modeled separately using, say, multinomial logistic regression. This approach is scalable to a large number of responses and high-dimensional predictors \citep{zhu2004classification,glmnet2,vincent2014sparse}, but entirely ignores the dependence between response variables. 
    On the other extreme, one could define a univariate categorical response variable based on the set of all possible category combinations and methods designed for a univariate categorical response could be applied. This is the regression analog of modeling counts in a $M$-way contigency table as a multinomial random variable \citep[Section 1.2]{agresti1992survey}. For example, in applications with only binary responses, this would require treating $M$ binary response variables as a univariate categorical variable with $2^M$ categories. In Section \ref{sec:real}, we consider a genomic application with $M=14$ binary response variables, so this approach would lead to an unwieldy $2^M=16384$ categories. This approach allows for arbitrary dependence among responses, but in so doing, treats $\mbP(\mbx)$ as a vector and thus fails to exploit its special tensor structure. Moreover, this approach would require an enormous amount of data for model fitting. If even a single category combination is not observed in the training data, this approach cannot be applied directly. Finally, when the number of categorical responses is even moderately large, model interpretation will be difficult because the number of parameters grows exponentially with the number of responses. 

For the problem of modeling the conditional probability tensor function $\mbP$, we refer to these two approaches as \textit{separate} modeling (of each response) and \emph{vectorized} modeling (of the combined-category response), respectively.
%The vectorized approach transforms the multivariate response into a univariate categories-combined response and is thus equivalently to transforming $\mbP$ into a vector. 
The objective of this article is to propose an alternative to these two approaches; a method which can model complex dependencies among response variables like vectorized modeling, yet provides fitted models which can be computed and interpreted with the ease and scalability of separate modeling. Our approach exploits the connection between the conditional independence of $Y_1, \dots, Y_M$, or lack thereof, and the rank of the conditional probability tensor $\mbP(\mbx).$ Later, we prove that separate modeling implicitly assumes $\mbP(\mbx)$ is rank one, whereas vectorized modeling assumes no explicit upper bound on the rank of $\mbP(\mbx)$. We may thus characterize models for which $\mbP(\mbx)$ is low rank as intermediate to these two extremes. Neatly, we later show that both the separate model and vectorized model can be characterized as ``edge-cases" of a rank-constrained $\mbP(\mbx)$ when the rank is fixed at one or the rank is allowed to be as large as $\prod_{m=1}^M c_m/\max_{k \in [M]} c_k$, respectively. Section~\ref{sec:alternatives} contains comparisons of these approaches.

Motivated by this observation, in this article we propose a new method for multivariate categorical response which assumes the low-rankness of $\mbP(\mbx)$ {for all $\mbx.$} We show that
pursuing a low-rank decomposition of the conditional probability tensor $\mbP(\mbx)$ provides a natural, intuitive, and scalable way to model the 
complex dependencies among responses. However, because $\mbP(\mbx)$ consists of the probabilities from \eqref{ProbTensor}, 
there are intrinsic constraints (nonnegativity and ``sum-to-one'') not often encountered in standard tensor decomposition problems. 
%It is thus nontrivial to perform a low-rank tensor decomposition on $\mbP(\mbx)$ even just for a single $\mbx$. 
{To handle these difficulties, we assume that the conditional probability tensor function $\mbP$ can be decomposed into the weighted sum of rank one probability tensor functions. This assumption naturally allows $\mbP$ to be characterized as a mixture of regressions model, and implies the  low-rankness of $\mbP(\mbx)$ for all $\mbx \in \mathbb{R}^p.$ Moreover, by exploiting a latent variable interpretation of the probability tensor function, we can fit our model using penalized maximum likelihood, which can accommodate large $p$, large $M$, and large $c_m$.}

Before formally describing our proposed method and model, we first situate our work among existing methods for fitting \eqref{ProbTensor}.

In the statistical literature on categorical data analysis, existing methods for multivariate categorical response regression are primarily focused on developing parametric links between predictors and multiple categorical responses that allow for model interpretation in terms of marginal probabilities and higher-order associations \citep{molenberghs1999marginal,glonek1996class,ekholm2000association,mccullagh1989generalized}. For example, one set of link functions correspond to log-linear models, and another to multivariate logistic models \citep{glonek1995multivariate}. Other approaches propose nonparametric regression functions and allow for some response-specific predictors \citep{gao2001smoothing}. In general, these works adopt the vectorized modeling approach and are often not feasible when $M\geq 3$ and $p$ is large, where associations among response variables are difficult to parameterize. 

More recently, in the high-dimensional regime, \citet{molstad2020likelihood} proposed a novel penalty to enforce linear restrictions on the regression coefficient tensor under the vectorized model with multinomial link. Their penalty can lead to fitted models that can be interpreted in terms of which predictors affect only the marginal distributions of responses, the log odds ratios, or neither. However, their method does not easily generalize to more than two response variables. Theoretically, the estimation error bound for their method scales exponentially with $M$ (more specifically, scales in $\prod_{m=1}^{M}c_m$) rather than linearly (such as $\sum_{m=1}^{M}c_m$) as does our method.

Of course, others have recognized the need for models and methods specifically designed for multivariate categorical response. One class of methods, based on the notion of binary relevance, comes from the literature on \emph{multi-label classification} \citep{tsoumakas2007multi} in machine learning. Many binary relevance methods fit separate univariate models, and thus fail to account for dependence in the multivariate response \citep{dembczynski2012label,montanes2014dependent,zhang2018binary}. One binary relevance approach that accounts for dependence uses {classifier chains} \citep{read2011classifier,senge2013rectifying}. This approach fits univariate categorical response regressions for $Y_1, Y_2, \dots, Y_M$ successively. For each univariate fit, the responses from previous fits are used as predictors in subsequent fits. For example, one would fit $Y_1 \mid \mbX$, then $Y_2 \mid \mbX, Y_1$, and so on. The fitted models are thus typically interpreted in terms of specific (univariate) conditional distributions for each response, rather than the joint distribution of interest. There is also an extensive literature on \emph{unsupervised} modeling for multivariate categorical data \citep{fienberg2000contingency,dunson2009nonparametric,bhattacharya2012simplex}, but this is not applicable to regression. Finally, we note that the problem of modeling $\mbP$ is not related to recent work on categorical data analysis focused on handling high-dimensional categorical predictors \citep[e.g.,][]{stokell2021modelling}.

Our model and method is related to---although fundamentally distinct from---existing research on tensor decompositions and regression. We give a very brief and selective survey in the following and refer the interested readers to \citet{kolda2009tensor} and \citet{bi2020tensors}.
First, the idea of jointly modeling multiple responses using a (regularized) tensor decomposition is conceptually similar to that in tensor response regression \citep[e.g.,][]{lizhang2017tensor}, where the response is typically a continuous-valued tensor. However, our study is fundamentally different due to the discrete nature of the response. For example, continuous-valued tensor decompositions \citep[e.g.,][]{sun2017provable} are not applicable since they may not result in a valid probability tensor. Secondly, our regression problem is also distinct from recent studies focused on binary or categorical tensor decompositions \citep[e.g.,][]{wang2020learning}. Extensions of these unsupervised learning methods to our context is nontrivial, especially in settings where the predictor is high-dimensional.  Notably, \citet{yang2016bayesian} also used the term ``conditional probability tensor'', though their focus was on estimating conditional probabilities of a categorical $Y$ on multivariate categorical predictor $\mbX$, which is fundamentally different from \eqref{ProbTensor}.

This paper has multiple contributions. 
First, we propose a general method for modeling $\mbP$ which allows practitioners to consider alternatives to the separate and vectorized modeling approaches. 
Crucially, unlike existing approaches which do not assume conditional independence, our method is scalable to large $p$, large $M$, and large $c_m$'s, without sacrificing flexibility or interpretability.  The scalability allows for a broad range of potential applications, and the interpretability---in terms of both predictors selected and the estimated rank of the conditional probability tensor---
allows practitioners to gain novel scientific insights.

Second, we introduce the notion of a \textit{functional} tensor rank decomposition: a type of decomposition applicable to tensor-valued functions. Loosely, this type of decomposition assumes that the range of a tensor-valued function is a restricted set of low-rank tensors. Though we focus on its application to conditional probability tensor functions, this approach could also be applied in more general tensor response regression problems.

Third, in order to accommodate different scenarios, we propose new penalties to achieve highly interpretable global and local variable selection in mixture of regression models. 
%Interestingly, the penalties instrically resolve the issue of the parameter non-identifiability.
We devise an efficient algorithm which we prove to produce a sequence of iterates that monotonically increase the penalized observed data log-likelihood. Statistical properties are also established to illustrate how, in an idealized setting, our method scales with respect to the number of responses, the number of categories per response, and the number of predictors.

\section{Model}\label{sec:model}

\subsection{Decomposition of the conditional probability tensor function}
For a positive integer $M$, an $M$-way tensor (also known as an $M$th order tensor) is an array object $\mbA\in\mathbb{R}^{p_1\times\cdots\times p_M}$ for positive integers $p_1,\dots,p_M$. 
%Each dimension or way of the tensor is also called a \emph{mode}, so $p_m$ is its dimension along the $m$-th mode. 
For example, a vector is a one-way tensor, and a matrix is a two-way tensor. 
The tensor rank decomposition, also known as the CANDECOMP/PARAFAC decomposition, is a generalization of the singular value decomposition for matrices \citep{hitchcock1927expression,carroll1970analysis}. 
A rank-one tensor $\mbA\in\mathbb{R}^{p_1\times\cdots\times p_M}$ can be written as the outer product of $M$ vectors: $\mbA = \mba^{(1)}\circ \cdots \circ\mba^{(M)}$, which  is defined element-wise as $\mbA_{i_1,\dots,i_M}= \mba^{(1)}_{i_1}\cdots \mba_{i_M}^{(M)}$ for all $i_m\in[p_m]$ and $m\in[M]$. In general, a rank-$R$ tensor ($R\geq2$) can be written as the sum of $R$ rank-one tensors, each formed as the outer product of vectors $\mba_{r}^{(m)},\ r\in[R],\ m\in[M]$: $\mbA = \sum_{r=1}^R \mba_{r}^{(1)}\circ \cdots \circ\mba_{r}^{(M)}$. A tensor is said to be rank-$R$ if it can be decomposed into $R$ rank-one tensors but not into $r$ rank-one tensors for any $r< R$.

{{
    The goal of this paper is to provide a statistical modeling framework for the $M$-th order tensor $\mbP(\mbx)\in\mbbR^{c_{1}\times\cdots\times c_{M}}$, whose $(j_{1},\dots,j_{M})$th element is the conditional probability function $P_{j_{1}\dots j_{M}}(\mbx)$ defined in \eqref{ProbTensor}. 
To that end, we first provide some characterizations of $\mbP(\mbx)$ and its rank-$R$ decomposition which motivate our modeling approach in Section~\ref{subsec:lvmodel}.

First, given $\mbx$, there is no distinction between a conditional probability tensor $\mbP(\mbx)$ and an unconditional probability tensor (i.e., a probability tensor which is not a function of predictors). Thus, many of the results in this section apply to unconditional probability tensors. We will explain shortly how these results motivate our model for the function $\mbP$.

Let $\Ptensor\subset\mbbR^{c_1\times\cdots\times c_M}$ denote the set of $M$-way valid probability tensors, i.e., the set of $M$-way tensors satisfying non-negativity and sum-to-one constraints. We define the \emph{probability tensor rank} based on the tensor rank decomposition restricted to the set $\Ptensor$.
\begin{definition}\label{def:ptrank} The probability tensor rank of $\mbA\in\Ptensor$ is the minimal number $R$ such that $\mbA$ can be expressed as the weighted sum of $R$ rank-one probability tensors, $\mbA = \sum_{r=1}^R\delta_r\mbA_r$ for some $\delta_r>0$ and $\mbA_r\in\Ptensor^{(1)}$, $r \in [R]$, where $\Ptensor^{(1)}$ denotes the set of $M$-way rank-one probability tensors.
\end{definition}

In Definition \ref{def:ptrank}, a rank-one probability tensor is defined in the usual sense, i.e., it can be formed as the outer product of vectors. However, the probability tensor rank $R$ is based on a more restrictive decomposition, in which the weights are positive and the tensors $\mbA_r$ are elements of $\Ptensor^{(1)}$.
These restrictions prompt meaningful statistical and probabilistic interpretation on the probability tensor decomposition, as we discuss later. Henceforth, we say a probability tensor is rank-$R$ if its probability tensor rank is $R$.
\begin{remark}
Because of the additional restrictions, a rank-$R$ probability tensor decomposition in Definition~\ref{def:ptrank} is also a valid rank-$R$ CP decomposition. The probability tensor rank is always no less than the usual tensor CP rank, similar to the fact that the nonnegative rank of a nonnegative matrix \citep{nonnegativerank} is no less than its usual matrix rank. 
\end{remark}
The following two propositions establish upper bounds on the probability tensor rank. %, which also serve as the upper bounds on the CP rank.
\begin{proposition}\label{pn:upper_rank}
For any given $\mbx$, $\mbP(\mbx)$ has probability tensor rank $R\leq \prod_{m=1}^Mc_m/\max_{k\in[M]}c_{k}$.
\end{proposition}
Proposition~\ref{pn:upper_rank} implies that the CP rank is also less than or equals to $\prod_{m=1}^Mc_m/\max_{k\in[M]}c_{k}$. This implication also extends Theorem 1 and Corollary 1 of \citet{dunson2009nonparametric} by establishing the upper bound on the rank $R$ while \citet{dunson2009nonparametric} showed the existence of the rank.
When $M=2$, Proposition~\ref{pn:upper_rank} implies that the singular value decomposition of $\mbP(\mbx)$, as a $c_1\times c_2$ matrix, holds for rank $R\leq \min(c_1,c_2)$ for any given $\mbx$. This well-known fact for the singular value decomposition of a matrix is thus extended to our decomposition with nonnegativity and sum-to-one constraints. The result of Proposition~\ref{pn:upper_rank} formalizes the arguments outlined in equation (6) of \citet{johndrow2017tensor}. 

The upper bound in Proposition \ref{pn:upper_rank} assumes nothing about the dependence among the $M$ responses. Indeed, if $\mbP(\mbx)$ has rank $R = \prod_{m=1}^M c_m/\max_{k\in[M]}c_{k}$, the responses can be arbitrarily dependent. Parsimonious dependence structures, in contrast, can imply a tighter upper bound on the rank of $\mbP(\mbx)$. We present one such example in the following proposition.
\begin{proposition}\label{pn:upper_rank2}
For a given $\mbx$, if the responses form $L$ mutually independent groups---indexed by sets $G_1, \dots, G_L$ where\ $\cup_{l=1}^L G_l = [M]$ and $G_k \cap G_k' = \emptyset$ for $k \neq k'$---then $\mbP(\mbx)$ has probability rank $R\leq \prod_{l=1}^L(\prod_{m\in G_l} c_m/\max_{k\in G_l}c_{k})$. \end{proposition}

Proposition \ref{pn:upper_rank2} suggests that the rank of $\mbP(\mbx)$ is related to the complexity of the dependence among $Y_1,\dots,Y_M$ given $\mbX=\mbx$. 
To demonstrate this point, consider the application in Section~\ref{sec:real} where $c_1=\cdots=c_M=2$, $M=14$. The generic upper bound from Proposition~\ref{pn:upper_rank} is $2^{13}=8192$. Instead, suppose that at a given $\mbx$, eight of the responses were independent of all others, and the other six formed two groups of three responses which are mutually independent, i.e., $|G_1| = 3, |G_2| = 3, |G_4| = \cdots =  |G_{10}| = 1$. In this case, Proposition~\ref{pn:upper_rank2} shows $R\leq 2^4=16$. In our application, we actually find that $R \approx 5$ yields the best results in terms of test set log-likelihood. \citet{johndrow2017tensor} also provide also upper bounds on the probability tensor rank as a function of the sparsity in the log-linear model characterizing the joint distribution of the responses (in an unconditional setting), further reinforcing that parsimonious dependence structures imply probability tensor rank restrictions.

Considering the most extreme case of Proposition \ref{pn:upper_rank2}, we have the following well-known result about rank-one probability tensors as an immediate corollary. 
\begin{corol}\label{cy:rank1}
For a given $\mbx$, $\mbP(\mbx)$ is rank one if and only if the responses are independent.
\end{corol}
Of course, a probability tensor $\mbP(\mbx)$ must have at least rank-one because rank-zero, which corresponds to the tensor of zeros, would not yield a valid probability tensor. 

Up to this point, our results have applied to $\mbP(\mbx)$ for a given $\mbx$. In full generality, $\mbP(\mbx)$ may could have a distinct decomposition for each $\mbx$, where $R$ also varies with $\mbx$, but allowing this degree of flexibility would make regression modeling impracticable. 
Instead, to achieve parsimony, it is reasonable to assume that $\mbP(\mbx)$ will have the same (low) probability tensor rank for every $\mbx$, and moreover, the components of their decomposition will have the same functional form. To see how one could put this assumption to use, we consider the rank-one decomposition of $\mbP(\mbx)$. Specifically, the following result establishes the connection between the rank-one conditional probability tensor and the conditional independence of the responses given the predictor. 
\begin{theorem}\label{thm:rankone}
For any given $\mbx$, if $\mbP(\mbx) \in \Ptensor$ is rank-one, or equivalently, can be decomposed
\begin{equation}\label{rank1probT}
\mbP(\mbx) = \mbp_{1}(\mbx)\circ\cdots\circ\mbp_{M}(\mbx),
\end{equation}
for some $\mbp_{m}(\mbx)\in\mbbR^{c_m}$, then \eqref{rank1probT} must be true for $\mbp_m(\mbx)=\{\Prob(Y_{m}=1\mid\mbX=\mbx),\ldots,\Prob(Y_{m}=c_m\mid\mbX=\mbx)\}^\top.$ Consequently, if $\mbP(\mbx)$ is rank-one for all $\mbx$, then $Y_1,\dots,Y_M$ are conditionally independent given $\mbX$, and vice versa. 
\end{theorem}
}
The tensor rank-one decomposition in \eqref{rank1probT} is not unique, as we impose no positivity or length constraints on $\mbp_m(\mbx)\in\mbbR^{c_m}$. Theorem~\ref{thm:rankone} shows that a rank-one probability tensor can be decomposed into marginal probability vectors without loss of generality. This theorem thus gives a constructive and identifiable formulation of rank one probability tensor decomposition. As shown in Corollary~\ref{cy:rank1}, rank-one probability tensor \emph{for a given $\mbx$} is equivalent to independence of responses conditional on the event $\mbX=\mbx$. As shown in Theorem~\ref{thm:rankone}, rank-one probability tensor \emph{for all $\mbx$} is equivalent to conditional independence of responses given the random variable $\mbX$. 
The result of Theorem~\ref{thm:rankone} suggests that if, for example, we assumed the responses were conditionally independent given $\mbX$, then we are equivalently assuming that $\mbP(\hspace{2pt}\cdot\hspace{2pt}) = \mbp_1(\hspace{2pt}\cdot\hspace{2pt}) \circ \cdots \circ \mbp_M(\hspace{2pt}\cdot\hspace{2pt})$ for functions $\mbp_1, \dots, \mbp_M$ such that $\mbp_m:\mathbb{R}^p \to \Delta^{c_m - 1}$ where $\Delta^{c_m - 1} = \{\mbu \in \mathbb{R}^{c_m}: \mbu^\top \mathbf{1}_{c_m} = 1,  \mbu_k \geq 0 \text{ for all } k \in [c_m]\}$. 
Under this assumption, we could thus model each $\mbp_m$ using standard regression models for $Y_m$ on $\mbX$ separately. 
On the other hand, the low-rankness of $\mbP(\mbx)$ for a specific value $\mbx$ would not reduce the population model complexity. 

Thus, motivated by model parsimony, we introduce the \textit{functional rank} of the probability tensor function $\mbP:\ \mbbR^p \to\Ptensor$ defining the conditional probability mass \eqref{ProbTensor}.
\begin{definition}\label{def:fptrank} The functional rank of the probability tensor function $\mbP(\cdot)$ is the minimal number $R$ such that $\mbP(\cdot) = \sum_{r=1}^R\delta_r\mbP_r(\cdot)$ for $\delta_r>0$ and $\mbP_r:\ \mbbR^p\to\Ptensor^{(1)}$, $r\in[R]$. 
\end{definition}
From the above definition, $\mbP_r(\cdot)$ has functional rank one and $\mbP_r(\mbx)\in\Ptensor^{(1)}$ is a rank-one probability tensor for all $\mbx$.
We now consider generalizing the rank-one structure to an arbitrary rank-$R$.
Motivated by Theorem~\ref{thm:rankone} and Definition~\ref{def:fptrank}, the next theorem considers the decomposition of $\mbP(\cdot)$ into a sum of $R$ rank-one probability tensor functions (see Figure \ref{fig:decomp_illustration}). %Similar to \eqref{rank1probT}, the unconstrained vectors $\mbp_{mr}(\mbx)\in\mbbR^{c_m}$ in the tensor rank decomposition will be restricted to 
%$\Delta^{c_m - 1}$, without loss of generality.%the set of $c_m$-dimensional vectors with nonnegative elements, denoted by $\mbbR^{c_m}_{+}$, without loss of generality.
\begin{theorem}\label{thm:rankR}
If the functional rank of $\mbP:\ \mbbR^p\to\Ptensor$ is $R\geq1$, then it can be written as
\begin{equation}\label{rankRprobT}
\mbP(\cdot) = \sum_{r=1}^R\delta_r\mbP_r(\cdot)=\sum_{r=1}^R\delta_r\mbp_{1r}(\cdot)\circ\cdots\circ\mbp_{Mr}(\cdot),
\end{equation}
where  $\delta_r>0$, $\sum_{r=1}^R\delta_r=1$ and $\mbp_{mr}:\ \mbbR^p\mapsto \Delta^{c_m - 1}$ for $(m,r) \in[M] \times [R]$. Moreover, if the decomposition of the function $\mbP$ in \eqref{rankRprobT} holds, then there exists a categorical random variable $Z$ independent of $\mbX$ such that $\Prob(Z=r)=\delta_r$ and $\mbp_{mr}(\mbx)=\{\Prob(Y_{m}=1\mid\mbX=\mbx,Z=r),\ldots,\Prob(Y_{m}=c_m\mid\mbX=\mbx,Z=r)\}^\top$ for $r \in [R].$ Consequently, \eqref{rankRprobT} is equivalent to the conditional independence of $Y_1,\dots,Y_M$ given $(\mbX,Z)$.
%$\delta_r\in\mbbR$, and $\mbp_{mr}(\mbx)\in\mbbR^{c_m}$. There exists $\delta_r>0$ and $\mbp_{mr}(\mbx)\in\mbbR^{c_m}_{+}$  with $\sum_{r=1}^R\delta_r=1$  and $\mbp_{mr}(\mbx) \in \Delta^{c_m - 1}$ for $(m,r) \in[M] \times [R]$ such that  \eqref{rankRprobT} holds. 
\end{theorem}
Theorem~\ref{thm:rankR} shows that a conditional probability tensor with functional rank-$R$ can always be decomposed with additional constraints that $\sum_r\delta_r=1$ and $\mbp_{mr}:\ \mbbR^p\to\Delta^{c_m-1}$, without loss of generality. Recall that $\Delta^{c_m-1}$ represents the sets of valid probability mass functions of $c_m$-categorical random variable. Then the rank-$R$ decomposition can be interpreted as the product of marginal probabilities $\Prob(Z=r)=\delta_r$ and conditional probability tensors $\mbP_r(\mbx)$, which is defined as the probability function of $\mbY\mid(\mbX=\mbx,Z=r)$. When \eqref{rankRprobT} holds, such a categorial variable $Z$ always exists by this construction. The rank-one probability tensor $\mbP_{r}(\mbx)$ can then be decomposed into the product of $\mbp_{mr}(\mbx)=\{\Prob(Y_{m}=1\mid\mbX=\mbx,Z=r),\ldots,\Prob(Y_{m}=c_m\mid\mbX=\mbx,Z=r)\}^\top$ similar to the decomposition in Theorem~\ref{thm:rankone}. The rank-$R$ decomposition in \eqref{rankRprobT} is the key modeling assumption of our approach. The result of Theorem \ref{thm:rankR} suggests a natural population-level decomposition of $\mbP(\cdot)$ as illustrated in Figure \ref{fig:decomp_illustration} for $R=3$.

\newcommand{\Depth}{2}
\newcommand{\Height}{2}
\newcommand{\Width}{2}

\newcommand{\sDepth}{0.5}
\newcommand{\sHeight}{0.5}
\newcommand{\sWidth}{0.5}

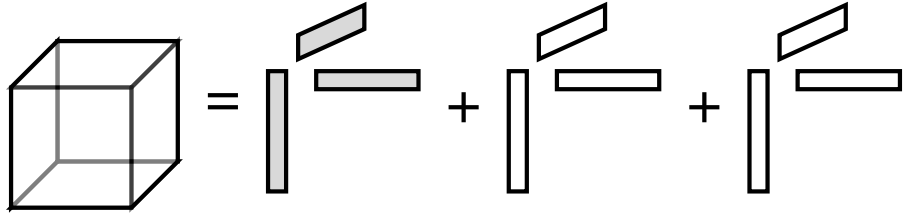
\begin{figure}
\centering
\hspace{0.15in}
\begin{tikzpicture}[scale=0.8]
% Large cube
\coordinate (O) at (0,0,0);
\coordinate (A) at (0,\Width,0);
\coordinate (B) at (0,\Width,\Height);
\coordinate (C) at (0,0,\Height);
\coordinate (D) at (\Depth,0,0);
\coordinate (E) at (\Depth,\Width,0);
\coordinate (F) at (\Depth,\Width,\Height);
\coordinate (G) at (\Depth,0,\Height);
\draw[ultra thick,black,fill=white!30] (O) -- (C) -- (G) -- (D) -- cycle;% Bottom Face
\draw[ultra thick,black,fill=white!30] (O) -- (A) -- (E) -- (D) -- cycle;% Back Face
\draw[ultra thick,black,fill=white!30] (O) -- (A) -- (B) -- (C) -- cycle;% Left Face
\draw[ultra thick,black,fill=white!30,opacity=0.5] (D) -- (E) -- (F) -- (G) -- cycle;% Right Face
\draw[ultra thick,black,fill=white!30,opacity=0.5] (C) -- (B) -- (F) -- (G) -- cycle;% Front Face
\draw[ultra thick,black,fill=white!30,opacity=0.5] (A) -- (B) -- (F) -- (E) -- cycle;% Top Face

% Equal sign
\coordinate (I) at (2.5,0.9,0);
\coordinate (J) at (3,0.9,0);
\coordinate (K) at (2.5,1.1,0);
\coordinate (L) at (3,1.1,0);
\draw[ultra thick,black] (I) -- (J); % Equal sign(lower line)
\draw[ultra thick,black] (K) -- (L); % Equal sign(upper line)

% Left Panel
\coordinate (M) at (3.5,-0.5);
\coordinate (N) at (3.8,-0.5);
\coordinate (P) at (3.8,1.5);
\coordinate (Q) at (3.5,1.5);
\draw[ultra thick,black,fill=gray!30] (M) -- (N) -- (P) -- (Q) -- cycle;

% Upper Panel
\coordinate (uM) at (4.0,1.7);
\coordinate (uN) at (5.1,2.2);
\coordinate (uP) at (5.1,2.6);
\coordinate (uQ) at (4.0,2.1);
\draw[ultra thick,black,fill=gray!30] (uM) -- (uN) -- (uP) -- (uQ) -- cycle;

% Right Panel
\coordinate (rM) at (4.3,1.2);
\coordinate (rN) at (6,1.2);
\coordinate (rP) at (6,1.5);
\coordinate (rQ) at (4.3,1.5);
\draw[ultra thick,black,fill=gray!30] (rM) -- (rN) -- (rP) -- (rQ) -- cycle;

% Plus sign
\coordinate (I) at (6.5,0.9,0);
\coordinate (J) at (7,0.9,0);
\coordinate (K) at (6.75,0.65,0);
\coordinate (L) at (6.75,1.15,0);
\draw[ultra thick,black] (I) -- (J); % Equal sign(lower line)
\draw[ultra thick,black] (K) -- (L); % Equal sign(upper line)

% Left Panel
\coordinate (M) at (7.5,-0.5);
\coordinate (N) at (7.8,-0.5);
\coordinate (P) at (7.8,1.5);
\coordinate (Q) at (7.5,1.5);
\draw[ultra thick,black,fill=white!30] (M) -- (N) -- (P) -- (Q) -- cycle;

% Upper Panel
\coordinate (uM) at (8.0,1.7);
\coordinate (uN) at (9.1,2.2);
\coordinate (uP) at (9.1,2.6);
\coordinate (uQ) at (8.0,2.1);
\draw[ultra thick,black,fill=white!30] (uM) -- (uN) -- (uP) -- (uQ) -- cycle;

% Right Panel
\coordinate (rM) at (8.3,1.2);
\coordinate (rN) at (10,1.2);
\coordinate (rP) at (10,1.5);
\coordinate (rQ) at (8.3,1.5);
\draw[ultra thick,black,fill=white!30] (rM) -- (rN) -- (rP) -- (rQ) -- cycle;

% Plus sign
\coordinate (I) at (10.5,0.9,0);
\coordinate (J) at (11,0.9,0);
\coordinate (K) at (10.75,0.65,0);
\coordinate (L) at (10.75,1.15,0);
\draw[ultra thick,black] (I) -- (J); % Equal sign(lower line)
\draw[ultra thick,black] (K) -- (L); % Equal sign(upper line)

% Left Panel
\coordinate (M) at (11.5,-0.5);
\coordinate (N) at (11.8,-0.5);
\coordinate (P) at (11.8,1.5);
\coordinate (Q) at (11.5,1.5);
\draw[ultra thick,black,fill=white!30] (M) -- (N) -- (P) -- (Q) -- cycle;

% Upper Panel
\coordinate (uM) at (12.0,1.7);
\coordinate (uN) at (13.1,2.2);
\coordinate (uP) at (13.1,2.6);
\coordinate (uQ) at (12.0,2.1);
\draw[ultra thick,black,fill=white!30] (uM) -- (uN) -- (uP) -- (uQ) -- cycle;

% Right Panel
\coordinate (rM) at (12.3,1.2);
\coordinate (rN) at (14,1.2);
\coordinate (rP) at (14,1.5);
\coordinate (rQ) at (12.3,1.5);
\draw[ultra thick,black,fill=white!30] (rM) -- (rN) -- (rP) -- (rQ) -- cycle;

\end{tikzpicture}   
\caption{\label{fig:decomp_illustration} Diagram of the decomposition of the probability tensor $\mbP(\mbx)$ (white cube) into the sum of $R=3$ rank-one tensors. The gray rank-one tensor represents, for example, $\delta_1\mbp_{11}(\mbx)\circ\cdots\circ\mbp_{M1}(\mbx)$ from the decomposition in \eqref{rankRprobT}.}
\end{figure}

\subsection{Finite mixture of regressions model}\label{subsec:lvmodel}

Recall that for a $\mbP$ with rank-$R$ functional probability tensor decomposition \eqref{rankRprobT}, there exists latent categorical variable $Z\in [R] = \{1,\dots,R\}$ independent of $\mbX$ such that $Y_1,\dots,Y_M$ are conditionally independent given $\mbX$ and $Z$. 
Specifically, we have $\delta_r=\Prob(Z=r\mid\mbX=\mbx)=\Prob(Z=r)$ and $\mbp_{mr}(\mbx)=\Prob(Y_{m}=j_{m}\mid\mbX=\mbx,Z=r)$ such that \eqref{rankRprobT} is satisfied in Theorem~\ref{thm:rankR}. It follows that
\begin{equation} \label{bayes}
P_{j_{1}\dots j_{M}}(\mbx) = \sum_{r=1}^R \big\{ \Prob(Z=r) \prod_{m=1}^M\Prob(Y_{m}=j_{m}\mid\mbX=\mbx,Z=r)\big\}.
\end{equation}
This connection between \eqref{rankRprobT} and \eqref{bayes} naturally leads to a finite mixture of regression model, which implies the population level tensor decomposition \eqref{rankRprobT} for all values of $\mbx$ by assuming a parametric link for the conditional probability function $\Prob(Y_{m}=j_{m}\mid\mbX=\mbx,Z=r)$ for $(m,r) \in[M] \times [R]$. 
With this latent categorical variable $Z\in[R]$, our proposed model is thus
\vspace{-10pt}
\begin{equation}\label{LatentMixtureModel}
\Prob(Z=r) = \delta_r,\quad f(\mbY\mid \mbX=\mbx, Z=r, \boltheta) =  \prod_{m=1}^{M}f_r(Y_m\mid \mbX=\mbx, \bolbeta_{mr}),
\end{equation}
where $f$ and $f_r$ denote generic probability mass functions and $\bolbeta_{mr}$ (to be specified later) is the model parameter for the regression of $Y_m$ on $\mbX$ given $Z=r$. The parameters in this model are denoted $\boltheta = \{\delta_r,\boltheta_r\}_{r\in[R]} = \{\delta_r,\bolbeta_{1r},\dots, \bolbeta_{Mr}\}_{r\in[R]}$. 
Of course, we can write the joint probability mass function of interest as a mixture of regressions without $Z$ as
\begin{equation}\label{MixtureModel}
f(\mbY\mid \mbX=\mbx,\boltheta) = \sum_{r=1}^R\delta_r f_r(\mbY\mid\mbX=\mbx,\boltheta_r) = \sum_{r=1}^R\big\{ \delta_r \prod_{m=1}^{M}f_r(Y_m\mid \mbX=\mbx, \bolbeta_{mr})\big\}.
\end{equation}

The model \eqref{LatentMixtureModel} has an intuitive interpretation: there are $R$ latent states indexed by $Z$, and conditional on the latent state and $\mbX$, the response variables are independent. For each value of the latent variable $Z$, we have a distinct sub-model with conditionally independent responses (i.e., separate models). Each category of $Z$ corresponds to a rank-one probability tensor (e.g., the gray rank-one tensor from Figure \ref{fig:decomp_illustration}). 
The vectors $\mbp_{mr}(\mbx)$ in \eqref{rankRprobT} naturally consist of probabilities $\Prob(Y_m=1\mid Z=r, \mbX=\mbx), \dots, \Prob(Y_m=c_m\mid Z=r, \mbX=\mbx)$.
Notably, maximum likelihood estimates of $\boltheta$, and consequently, of $\mbP$, can be obtained even when all category combinations are not observed in the training data. This is because each set of parameters $\boltheta_r$ corresponds to marginal conditional probabilities for each response.

For the remainder, we assume a multinomial logistic regression model for each $Y_m \mid (\mbX = \mbx, Z = r)$. The regression coefficients $\bolbeta_{mr}\in\mbbR^{p\times c_m}$ thus characterize the $c_m$ possible outcomes of a multinomial random variable $Y_m$ (based on a single trial), whose mass function $f_r(Y_m \mid \mbX = \mbx, \bolbeta_{mr})$ consists of the probabilities 
\begin{equation}\label{multinomial}
\Prob(Y_m = j \mid \mbX=\mbx, \bolbeta_{mr}) = \frac{\exp(\bolbeta_{mrj}^\top \mbx)}{\sum_{k=1}^{c_m}\exp(\bolbeta_{mrk}^\top \mbx)}, ~~~~ j\in[c_m],
\end{equation}
where $\bolbeta_{mrj}\in\mbbR^p$ is the $j$th column of $\bolbeta_{mr}\in\mbbR^{p\times c_m}$. Under \eqref{multinomial}, the $\bolbeta_{mr}$ are not identifiable. This can be resolved by imposing the condition $\sum_{j=1}^{c_m} \bolbeta_{mrj} = 0$, a ``sum-to-zero'' constraint. Under this constraint, the model formulation and parameterization lead to identifiable parameters and conditional distributions.  Later, we explain that when using our proposed penalties, this constraint is enforced automatically. 

Our functional probability tensor decomposition is related to the latent class model, which has been extensively studied and applied in psychological and epidemiological research. In latent class models, multiple binary or categorical variables ($\mbY$) are measured as surrogates for estimation and inference on the unobservable definitive categorical outcome ($Z$), which characterizes underlying population heterogeneity and the subjects' class memberships. In particular, \citet{bandeen1997latent}, \citet{huang2004building}, and \citet{ouyang2022identifiability}---among others---consider the regression extension of latent class models by including covariates (e.g., $\mbX$) and modeling $Z$ given $\mbX$, and $\mbY$ given $Z$ (or more generally, $\mbY$ given $Z, \mbX$) with generalized linear models.  Based on this connection, our work provides a new perspective on latent class models with covariates, and a means for application thereof with high-dimensional predictors (Section \ref{sec:estimation}). 

Mathematically, our model can be transformed into the regression-based latent class model of \citet{huang2004building} by replacing $\delta_r=\Prob(Z=r)$ with $\delta_r(\mbx)=\Prob(Z=r\mid\mbX=\mbx)$. We do not purse this direction for multiple reasons.  First, there is a philosophical difference between the latent class model and the multivariate categorical response regression model. Allowing $Z$ (instead of, or in addition to, $\mbY$) to be dependent on $\mbX$ is crucial in latent class models because $Z$ is the definitive outcome and $\mbY$ is a surrogate thereof. For example, \citet{bandeen1997latent} assume $\mbY$ is independent of $\mbX$ given the latent variable $Z$.  However, this is unnecessary in our context where the outcome of interest is $\mbY$ and $Z$ is only used to introduce dependence among the $Y_m$. The fundamentally different applications lead to different model assumptions.

Secondly, assuming $Z$ to be independent of $\mbX$ simplifies the model interpretation. As seen in Figure~2, our latent variable model can be viewed as a mixture of generalized linear regressions, where $\delta_r=\Prob(Z=r)=\Prob(Z=r\mid \mbX)$, $r\in[R]$, are non-stochastic weights of the $R$ mixtures. The assumption of non-stochastic weights is widely adopted in the study of the finite mixture of regression (FMR) model; see  \citet{khalili2007variable} for a formal definition on the parameter space of the FMR model. Additionally, our assumptions of independence between $Z$ and $\mbX$ helps parameter identifiability, a fundamental issue in latent class models \citep{ouyang2022identifiability}.

Finally, we note that it is relatively straightforward to generalize our method by modeling $\delta_r(\mbx)$ as a multinomial logistic regression of $Z\mid \mbX$. This extension is almost identical to the extension from the FMR model to the mixture of experts model \citep{jacobs1991adaptive}. However, because \eqref{MixtureModel} is already very flexible due to the large number of regression parameters, the extension from $\delta_r$ to $\delta_r(\mbx)$ sometimes offers little improvement in predicting $\mbY$. We conducted a simulation example to illustrate this point: see our discussion thereof in Section \ref{sec:sim}.

 \subsection{Comparison with alternative approaches}\label{sec:alternatives}
 In this section, we compare our model assumptions \eqref{MixtureModel} and \eqref{multinomial} to the two other approaches for fitting $M$ categorical responses $\mbY=(Y_1,\dots,Y_M)$ on the predictor $\mbX\in\mbbR^p$. For the sake of comparison, we use multinomial logistic links for all approaches.

 The first, and most naive, direct approach is separate modeling. This model assumes that 
 \begin{equation}\label{sep_model}
 \Prob(Y_m = j \mid \mbX=\mbx) = \frac{\exp(\boleta_{m j}^\top\mbx)}{\sum_{k=1}^{c_m}\exp(\boleta_{m k}^\top\mbx)},\quad j \in[c_m],\ m\in[M],
 \end{equation}
 where $\boleta_{mj}\in\mbbR^p$ for $j \in[c_m]$ and $m \in [M]$. This model is equivalent to our model with $R=1$. If \eqref{sep_model} is true, then our model with $R>1$ becomes over-parameterized but can still provide consistent estimates of $\boleta_{mj}$'s (with some asymptotic efficiency loss).  If \eqref{MixtureModel} and \eqref{multinomial} are true for $R>1$, then separate fitting based on \eqref{sep_model} will not only lose the interrelationship between responses, but will  be incorrect for the marginal probabilities $\Prob(Y_m\mid\mbX=\mbx)$,  $m\in[M]$. This may be somewhat surprising, but can be seen from the latent variable representation
 $$\Prob(Y_m=j\mid\mbX=\mbx)=\sum_{r=1}^R\delta_r\Prob(Y_m=j\mid\mbX=\mbx,Z=r) = \sum_{r=1}^R\delta_r\dfrac{\exp(\bolbeta_{mrj}^\top\mbx)}{\sum_{k=1}^{c_m}\exp(\bolbeta_{mrk}^\top\mbx)},$$ which can not be rewritten as proportional to $\exp(\boleta_{mj}^\top\mbx)$. Intuitively, the latent variable $Z$ introduces heterogeneity, and hence nonlinearity, in the conditional probability function $\Prob(Y_m=j\mid\mbX=\mbx)$. As a result, separate model fitting is insufficient even when there is only one response variable. This is analogous to fitting a linear model to a mixture of regressions with heterogeneous sub-populations.

 The second direct approach is vectorized modeling. As mentioned, this approach transforms $\mbY$ into a univariate categorical response $\boldsymbol{Y}_*$ with $c_* = \prod_{m=1}^Mc_m$ categories. The corresponding multinomial logistic regression model assumes 
 \begin{equation}\label{joint_model}
 \Prob(\boldsymbol{Y}_* = j\mid \mbX=\mbx) = \frac{\exp(\bolgamma_{j}^\top\mbx)}{\sum_{k=1}^{c_*}\exp(\bolgamma_{k}^\top\mbx)},\quad j\in[c_*], 
 \end{equation}
 where $\bolgamma_j\in\mbbR^p$ for $j\in[c_*]$. For example, \citet{molstad2020likelihood} assume \eqref{joint_model} and impose linear restrictions on the matrix of $\bolgamma_{j}$'s.  Similar to the separate fitting approach, if \eqref{MixtureModel} and \eqref{multinomial} are true for $R>1$, then this joint fitting based on \eqref{joint_model} will also be incorrect even for marginal probabilities $\Prob(Y_m\mid\mbX=\mbx)$, $m\in[M]$. On the other hand, if \eqref{joint_model} is correct, then the rank $R$ in our model may be as large as $c_*$. That is, we have $c_*$ rank-one tensors that each consists of one element of the probability tensor $\mbP(\mbx)$. 
 The number of free parameters is $p(c_*-1)=p(\prod_m^M c_m-1)$ for \eqref{joint_model} and $(R-1)+p R\sum_{m}(c_m-1)$ for our latent variable model \eqref{MixtureModel} and \eqref{multinomial}. 
 For example, consider the scenario where $c= c_1=\cdots=c_M$. Then the number of free parameters becomes $p(c^M-1)$ for the vectorized model \eqref{joint_model} and $(R-1)+p R(c-1) M$ for the rank $R$ version of our model. 
 As the number of responses $M$ increases, the complexity of \eqref{joint_model} increases exponentially the order of $O(pc^M)$ while our model's complexity increases linearly in the order of $O(pcMR)$. As the number of categories for each response $c$ increases, our model's complexity still increases linearly in $c$ but the complexity of \eqref{joint_model} increases more rapidly as $c^M$ as $M\geq 2$. To gain further intuition, when $c=M=4$ and $p=100$ (as in our simulation studies), the joint model has $25500$ free parameters and our model has $2401$ when $R=2$ or $3602$ when $R=3$.  

 A practical advantage of our approach is that its estimation algorithm is much more scalable to large $c$, $M$, and $p$ than estimators based on the vectorized model \eqref{joint_model}. When $M$ is large, methods that fit \eqref{joint_model} become infeasible partly due to the enormous number of parameters.  We discuss our computational approach in the subsequent section.

\section{Penalized maximum likelihood estimation}\label{sec:estimation}
Let the observed data be $\{(\mbY_i,\mbx_i)\}_{i=1}^n$ where $\mbY_i = (Y_{1i},\dots, Y_{Mi})$  for $i \in [n] = \{1, \dots, n\}$. 
Recall that $\boltheta$ denotes all of the unknown parameters $\{(\delta_r, \bolbeta_{1r}, \dots, \bolbeta_{Mr})\}_{r\in[R]} \in \mathcal{D}^R$ where $\mathcal{D} = (0,1) \times \mathbb{R}^{p \times c_1}\times \cdots \times \mathbb{R}^{p \times c_M}$ with $\sum_{r=1}^R \delta_r = 1.$ The conditional log-likelihood of $\mbY \mid \mbX$ evaluated at $\boltheta$ is 
\begin{equation}\label{eq:observed_data}
\sum_{i=1}^n\log\left[\sum_{r=1}^R\delta_r\big\{\prod_{m=1}^Mf_r(Y_{mi}\mid\mbX_i= \mbx_i,\bolbeta_{mr})\big\} \right].
\end{equation}
In this section, we first describe the standard EM algorithm for maximizing \eqref{eq:observed_data} over $\mathcal{D}^R$. The standard EM algorithm, which iterates between the expectation (E) step and the maximization (M) step, is only applicable in the classical low-dimensional setting. To address settings with high-dimensional predictors, we later discuss how to maximize a penalized version of \eqref{eq:observed_data}. In particular, we devise a computational algorithm that replaces the penalized M-step with an approximation guaranteed to monotonically increase the objective function.%The regularized EM algorithm is computationally efficient and guaranteed to increase the penalized conditional log-likelihood monotonically.

\subsection{The EM algorithm and its parallel M-step}\label{subsec:estimator}
% Thanks to our conditionally independent responses in each latent state, we show in this section that the key maximization step in the EM algorithm is separable in terms of both responses and latent states. 

The standard EM algorithm will deal with the complete-data log-likelihood; that is, the log-likelihood of $(\mbY,Z)\mid\mbX$, treating $Z$ as if it were observable.
Let $Z_{ir} = \mathbf{1}(Z_i=r)$. Recalling that $Z$ is independent of $\mbX$ with $\Prob(Z=r\mid\mbX)=\Prob(Z=r)=\delta_r$, the log-likelihood of $(\mbY,Z)\mid\mbX$ evaluated at $\boltheta$ is thus
\begin{equation}\label{eq:lik_Z}
\mathcal{L}(\boltheta) 
= \sum_{i=1}^n \sum_{r=1}^R Z_{ir}\log\left\{f_r(\mbY_i\mid\mbX_i = \mbx_i,\boltheta_r) \right\} + \sum_{i=1}^n \sum_{r=1}^R Z_{ir}\log(\delta_r),
\end{equation}
where $f_r(\mbY_i \mid \mbX_i = \mbx_i,\boltheta_r) = \prod_{m=1}^Mf_r(Y_{mi}\mid \mbX_i= \mbx_i,\bolbeta_{mr})$ by definition. Each iteration of the EM algorithm, indexed by $t=0,1,2,\dots$, consists of two steps. In the E-step, we compute the $\mathcal{Q}$-function at $t$th iterate $\boltheta^{(t)}$, $
\mathcal{Q}(\boltheta\mid \boltheta^{(t)}) = \E\left[ \mathcal{L}(\boltheta) \mid \{(\mbY_i, \mbx_i)\}_{i=1}^n, \boltheta^{(t)} \right]$.
To do so, we first compute the conditional estimate of $\pi_{ir}= \E(Z_{ir})$, the probability the $Z_i = r$, given the observed data and $\boltheta^{(t)}$, 
\begin{equation}\label{Estep}
\pi_{ir}^{(t)} = \dfrac{ \delta_r^{(t)} f_r(\mbY_i\mid\mbX_i=\mbx_i,\boltheta_r^{(t)})}{\sum_{s=1}^R\delta_s^{(t)} f_s(\mbY_i\mid\mbX_i=\mbx_i,\boltheta_r^{(t)})}.
\end{equation}
Then we can express the $\mathcal{Q}$-function as
\begin{equation}\label{eq:q_function}
\mathcal{Q}(\boltheta\mid \boltheta^{(t)})  = \sum_{i=1}^n\sum_{r=1}^R \left[ \pi_{ir}^{(t)}\sum_{m=1}^M\log\{f_r(Y_{mi}\mid\mbX_i = \mbx_i,\bolbeta_{mr})\} + \pi^{(t)}_{ir} \log(\delta_r)\right].
\end{equation}
In the M-step, we compute $\boltheta^{(t+1)}$, which we define as the maximizer of \eqref{eq:q_function} with respect to $\boltheta$. 
One can verify that $\delta_r^{(t+1)} = n^{-1}\sum_{i=1}^n \pi_{ir}^{(t)}$, so the main challenge is maximizing the $\mathcal{Q}$-function with respect to the regression coefficients $\bolbeta_{mr}\in\mbbR^{p\times c_m}$ for $(m,r) \in[M] \times [R]$. 

From the first term in the $\mathcal{Q}$-function, one can see that the maximization with respect to the $\bolbeta_{mr}$ is separable across each $(m,r)$ combination. Therefore, for $(m,r) \in [M] \times [R]$ in an embarassingly parallel fashion, we need only compute 
\begin{equation}\label{obj_betamr}
\begin{split}
\bolbeta_{mr}^{(t+1)} & = \argmax_{\bolbeta_{mr}\in\mbbR^{p\times c_m}} \ell_{mr}(\bolbeta_{mr}\mid\boltheta^{(t)}), \\
& \ell_{mr}(\bolbeta_{mr}\mid\boltheta^{(t)})  = \sum_{i=1}^{n}\pi_{ir}^{(t)} \log\{f_{r}(Y_{mi}\mid\mbX_i = \mbx_i,\bolbeta_{mr})\}.
\end{split}
\end{equation}
The solution to the above optimization problem is obtained by fitting a weighted multinomial logistic regression model of $Y_m$ on $\mbX$. This could be done using a modified version of the standard computational approaches, e.g., a quasi-Newton algorithm. This special structure naturally lends itself to settings with large $M$ and large $R$.

The update \eqref{obj_betamr} reinforces the generality of our latent variable model. We could replace the assumption of multinomial logistic link in \eqref{multinomial} with a different assumption on $f_{r}(Y_{mi}\mid\mbX_i = \mbx_i,\bolbeta_{mr})$. The only necessary modification of the estimation procedure is in \eqref{obj_betamr}.

\subsection{Penalties on the regression coefficient tensor}\label{subsec:penalty}

To address the $p > n$ case, we propose to maximize a penalized version of \eqref{eq:observed_data}. Imposing penalties on each $\bolbeta_{mr}$ separately is possible, but may lead to fitted models which are difficult to interpret. Moreover, by imposing penalties across both mixture and response components, efficiency can be greatly improved. 
To achieve this, %we consider the joint estimation of all the regression coefficients and 
first organize all $\bolbeta_{mr}\in\mbbR^{p\times c_m}$ into a tensor parameter $\mbB \in \mathbb{R}^{p \times R \times C}$, where $C=\sum_{m=1}^M c_m$ and define the $m$th \emph{mode-2 slice} of $\mbB$ as $\mbB_{[:,m,:]}=(\bolbeta_{m1},\dots,\bolbeta_{mR})\in\mbbR^{p\times C}$ for $m \in [M]$. %where $\mbB_j \equiv ([\bolbeta_{11}]_{j,:}',[\bolbeta_{21}]_{j,:}', \dots, [\bolbeta_{RM}]_{j, :}')' \in \mathbb{R}^{R \sum_{k=1}^M c_k}$ denotes the $j$-th rows of all $\bolbeta_{mr}$'s, we would compute 
We propose to estimate the parameters $\boltheta$ using
\begin{equation}\label{eq:penalized_obs_loglik}
\argmax_{\boltheta \in \mathcal{D}^R} \left\{\mathcal{F}(\boltheta)  - \mathcal{P}_{\lambda}(\mbB) \right\},
\end{equation}
where $\mathcal{F}$ is the observed data conditional log-likelihood in \eqref{eq:observed_data} and $\mathcal{P}_\lambda $ is a sparsity-inducing penalty by the tuning parameter $\lambda>0$.
To compute \eqref{eq:penalized_obs_loglik}, we need only modify the M-step of the EM algorithm from Section \ref{subsec:estimator} to be replaced with the following joint optimization problem
\begin{equation}\label{Mstep}
% \argmax_{\mbB \in \mathbb{R}^{p \times R \times C}}  \left( \sum_{i=1}^n\sum_{r=1}^R \left[ \pi_{ir}^{(t)}\sum_{m=1}^M\log\{f_r(Y_{mi}\mid\mbX_i = \mbx_i,\bolbeta_{mr})\}\right] 
\argmax_{\mbB \in \mathbb{R}^{p \times R \times C}} \mathcal{M}_\lambda(\mbB\mid  \boltheta^{(t)}),~~~ \mathcal{M}_\lambda(\mbB\mid  \boltheta^{(t)}) =  \sum_{m=1}^M\sum_{r=1}^R \ell_{mr}(\bolbeta_{mr}\mid \boltheta^{(t)})  -  \mathcal{P}_\lambda(\mbB),
\end{equation}
where $\ell_{mr}(\cdot\mid\boltheta^{(t)})$ is defined in  \eqref{obj_betamr} and $\mathcal{P}_\lambda $ is a sparsity-inducing penalty with tuning parameter $\lambda>0$.
If $\lambda = 0$, \eqref{Mstep} would reduce to \eqref{obj_betamr}. However, when $\lambda > 0$, the penalized objective function $\mathcal{M}_\lambda(\mbB\mid  \boltheta^{(t)}) $ may not be separable across responses and mixture components depending on the choice of penalty $\mathcal{P}_\lambda$: we propose two such penalties which correspond to distinct types of variable selection.

\begin{figure}[t]
\centering
\includegraphics[scale=0.33]{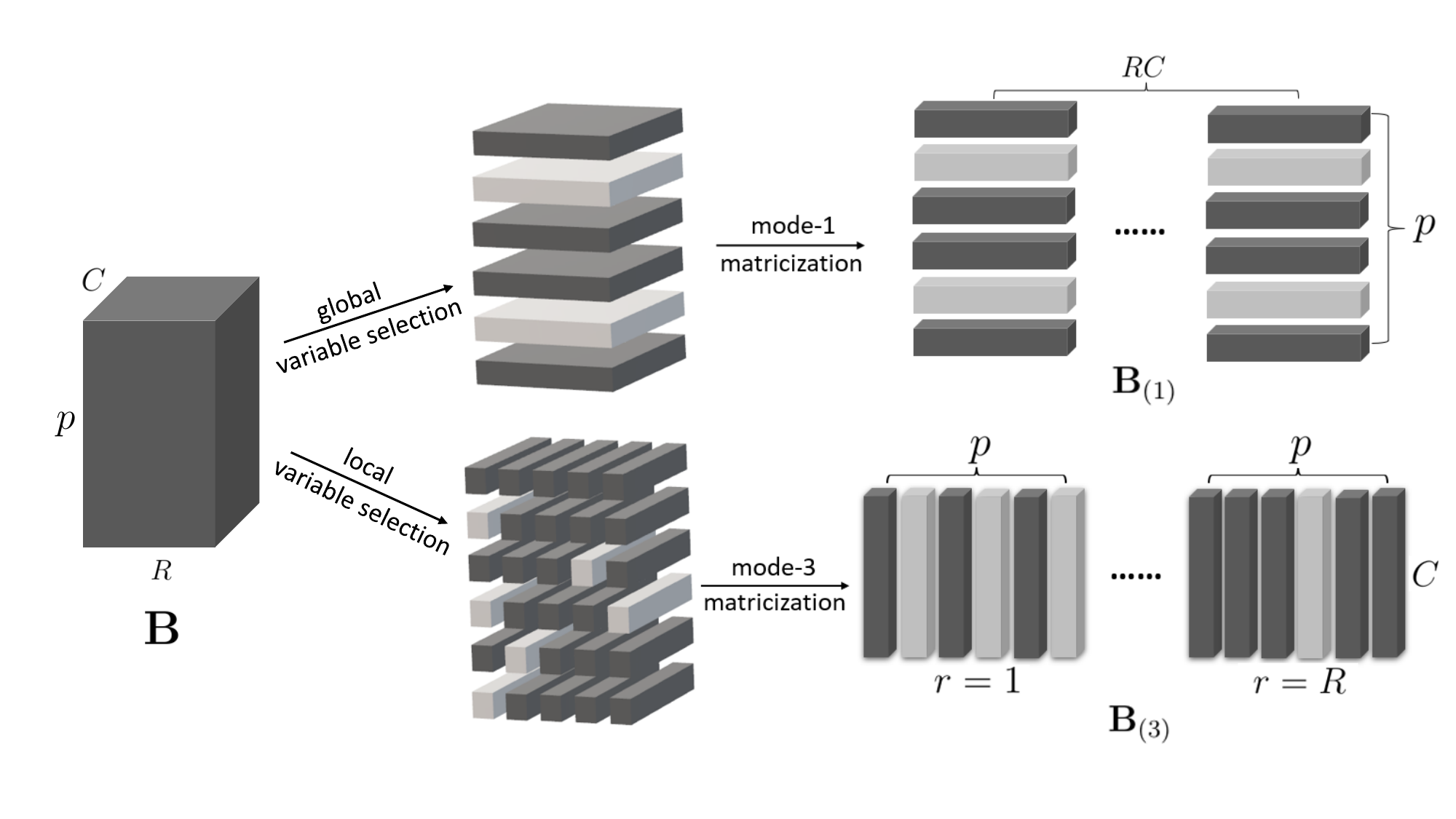}
\caption{Visualizing the global and local variable selections and the corresponding sparsity patterns in the tensor regression parameter $\mbB$. The two plots in middle respectively show the mode-1 slices and the mode-3 fibers to be selected. Our blockwise proximal gradient descent algorithm updates one row of $\mbB_{(1)}$ at a time for the penalty $\mathcal{G}_\lambda$, and updates one column of $\mbB_{(3)}$ at a time for the penalty  $\mathcal{H}_\lambda$. }\label{fig:penalty}
\end{figure}

First, we consider global variable selection. We say that the $j$th predictor is irrelevant if a change in the $j$th component of $\mbx$ does not change $\mbP(\mbx)$ for all $\mbx \in \mathbb{R}^p$. Under our model assumption on $\mbP$, for the $j$th variable to be irrelevant it must be that $\mbB_{[j,:,:]} =  (d_{j11} \cdot\mathbf{1}_{c_1}^\top, d_{j12} \cdot\mathbf{1}_{c_1}^\top, \dots, d_{jMR} \cdot\mathbf{1}_{c_M}^\top)^\top \in \mathbb{R}^{MC}$ for constants $d_{jmr} \in\mbbR$, $(m,r) \in [M] \times [R]$. Recall that we have over-parameterized $\mbB$ in the sense that each $\bolbeta_{mr}\in\mbbR^{p\times c_m}$ has only $(c_m-1)$ identifiable columns (see the ``sum-to-zero'' constraint under equation \eqref{multinomial}). This means that we may replace the $d_{jmr}$'s with $0$ without loss of generality and ensure parameter identifiability. This equivalence between predictor irrelevance and sparsity in $\mbB$ is discussed more in the theoretical analysis (Section~\ref{sec:theory}).  
%for all $\bolbeta_{mr} \in \mathbb{R}^{p \times C_m}$ with  $(m,r) \in [M]\times [R]$, $[\bolbeta_{mr}]_{j, :} = d \cdot 1_{C_m}$ for some constant $d \in \mathbb{R}$. To achieve this type of variable selection, we will require a generalization of the group lasso penalty which is imposed across latent states $Z$.
For such global variable selection, we propose the following penalty term as $\mathcal{P}_{\lambda}$ in \eqref{Mstep},
\begin{equation}\label{eq:penalty} 
\mathcal{G}_{\lambda}(\mbB) 
 = \lambda \sum_{j=1}^p \Vert\mbB_{[j,:,:]}\Vert_F = \lambda \sum_{j=1}^p \sqrt{\sum_{r=1}^R \sum_{m=1}^M \|[\bolbeta_{mr}]_{j,:}\|_2^2},
\end{equation} 
where $\lambda > 0$ is a user-specified tuning parameter and $[\bolbeta_{mr}]_{j,:}$ is the $j$th row of $\bolbeta_{mr}$. The penalty $\mathcal{G}_{\lambda}$ is nondifferentiable when for some $j \in [p] = \{1, \dots, p\}$, $[\bolbeta_{mr}]_{j,k} = 0$ for all $k \in [c_m]$ and $(m,r) \in [M] \times [R]$.  This penalty thus links the $\bolbeta_{mr}$ across both latent states and response variables. For large values of $\lambda$, this penalty will encourage estimates of $\mbP$ such that many predictors are estimated to be irrelevant by encouraging zeros across the same rows of all $RM$ coefficient matrices. 

Although $\mathcal{G}_\lambda$ can achieve a highly interpretable global form of variable selection, an alternative penalty, $\mathcal{H}_\lambda$, allows for variable selection specific to each latent state (i.e., local variable selection). Specifically, we also propose the penalty 
\begin{equation}\label{eq:alt_penalty} 
\mathcal{H}_\lambda (\boltheta) = \lambda \sum_{j=1}^p\sum_{r=1}^R \Vert\mbB_{[j,r,:]}\Vert_2 = \lambda \sum_{j=1}^p \sum_{r=1}^R \sqrt{\sum_{m=1}^M \|[\bolbeta_{mr}]_{j,:}\|_2^2}.
\end{equation} 
In contrast to $\mathcal{G}_\lambda$, the penalty $\mathcal{H}_\lambda$ assumes that for a particular value of the latent variable $Z$, a possibly unique set of predictors are important. This penalty allows practitioners to characterize the categories of latent variable $Z$ in terms of the predictors selected as relevant or not.  %As an convenient artifact, as long as $\lambda > 0$, both penalties naturally enforce the ``sum-to-zero'' constraint on the $\bolbeta_{mr}$. 

Figure~\ref{fig:penalty} provides a visualization of the global and local variable selection in terms of the sparsity of $\mbB$. The global penalty acts on entire mode-1 slices of the tensor parameter $\mbB$. As shown in the plot, the mode-1 matricization $\mbB_{(1)}\in\mbbR^{p\times RC}$ transforms each mode-1 slice into a row vector. The penalty $\mathcal{G}_\lambda(\mbB)$ is thus the group lasso penalty on rows of $\mbB_{(1)}$. On the other hand, the local penalty is targeting on a more refined sparsity pattern that is shown as mode-3 fibers of $\mbB$. Analogously, the mode-3 matricization $\mbB_{(3)}\in\mbbR^{C\times pR}$ aligns the fibers across $r\in[R]$. The penalty $\mathcal{H}_\lambda(\mbB)$ is thus the group lasso penalty on columns of $\mbB_{(3)}$.

For concreteness, we focus on computing \eqref{eq:penalized_obs_loglik} using the penalty $\mathcal{G}_\lambda$; only trivial modifications of our algorithm are needed to accommodate $\mathcal{H}_\lambda$. %In the following subsection, we detail exactly the steps of our proposed computational algorithm.

\subsection{Penalized EM algorithm}\label{subsec:ECM_alg}
In order to describe our approach, we express the objective function as 
$\mathcal{M}_\lambda(\mbB\mid  \boltheta^{(t)}) = \mathcal{M}_\lambda(\mbB_1, \mbB_2, \dots,\mbB_p\mid  \boltheta^{(t)})$ where $\mbB_j=\vecc(\mbB_{[j,:,:]})\in\mbbR^{RC}$ denotes the $j$th row of $\mbB_{(1)}$. 
 Similarly, we let $\mathcal{M}_{\lambda,j}(\cdot \mid \mbB_{-j}^{(t+1, t)}, \boltheta^{(t)}):\mathbb{R}^{RC} \to \mathbb{R}$ denote $\mathcal{M}_{\lambda}$ as a function of its $j$th argument alone with $\mbB_1, \dots, \mbB_{j-1}$ fixed at $\mbB_{1}^{(t+1)}, \dots, \mbB_{j-1}^{(t+1)}$ and $\mbB_{j+1}, \dots, \mbB_{p}$ at $\mbB_{j+1}^{(t)}, \dots, \mbB_{p}^{(t)}$. By construction, $\mathcal{M}_{0,j}(\cdot \mid \mbB_{-j}^{(t+1, t)}, \boltheta^{(t)})$ is simply the unpenalized version of $\mathcal{M}_{\lambda,j}(\cdot \mid \mbB_{-j}^{(t+1, t)}, \boltheta^{(t)})$.

Our approach is to update each $\mbB_j$ by maximizing a penalized quadratic minorizer \citep{lange2016mm} of $\mathcal{M}_{0,j}(\cdot \mid \mbB_{-j}^{(t+1,t)}, \boltheta^{(t)})$ constructed at $\mbB_j^{(t)}$ with all other $\mbB_{k}$ for $k \neq j$ fixed at their current iterates. Our minorizing objective function for $\mathcal{M}_{\lambda,j}(\mbb \mid \mbB_{-j}^{(t+1, t)}, \boltheta^{(t)})$, where $\mbb\in\mbbR^{RC}$, is 
\begin{eqnarray}
\nonumber
\widetilde{\mathcal{M}}_{\lambda,j}(\mbb \mid \mbB_{-j}^{(t+1,t)}, \boltheta^{(t)})  &=& \mathcal{M}_{0,j}(\mbB_{j}^{(t)} \mid \mbB_{-j}^{(t+1,t)}, \boltheta^{(t)}) - (2 \tau_j)^{-1}\|\mbb - \mbB^{(t)}_j\|_2^2 - \lambda \|\mbb\|_2 \\
&& ~~~~~+ \mathrm{tr}\big\{\nabla \mathcal{M}_{0,j}(\mbB^{(t)}_j \mid \mbB_{-j}^{(t+1,t)}, \boltheta^{(t)})^\top(\mbb - \mbB^{(t)}_j)\big\},\label{MMfunction}
\end{eqnarray}
for which, with a sufficiently small step size $\tau_j > 0$, we have the following result. 
\begin{theorem}\label{prop:descent}
Let $\tilde{\mbx}_j = (\mbx_{1j}, \dots, \mbx_{nj})^\top \in \mathbb{R}^n$ where $\mbx_{ij}$ is the $j$th component $\mbx_i$. For any fixed $0 < \tau_j \leq \{RM \|\tilde{\mbx}_j\|_2 (\max_{k\in[M]} \sqrt{c_k})\}^{-1}$, we have 
$ \widetilde{\mathcal{M}}_{\lambda,j}(\mbb \mid \mbB_{-j}^{(t+1,t)}, \boltheta^{(t)}) \geq \mathcal{M}_{\lambda,j}(\mbB_{j}^{(t)} \mid \mbB_{-j}^{(t+1,t)}, \boltheta^{(t)})$
for all $\mbb \in \mathbb{R}^{RC}$. Thus, if we define 
$\mbB_j^{(t+1)}=\argmax_{\mbb\in\mbbR^{RC}}\widetilde{\mathcal{M}}_{\lambda,j}(\mbb\mid \mbB_{-j}^{(t+1,t)}, \boltheta^{(t)}),$
then we are guaranteed that $\mathcal{M}_{\lambda,j}(\mbB_{j}^{(t+1)} \mid \mbB_{-j}^{(t+1,t)}, \boltheta^{(t)}) \geq \mathcal{M}_{\lambda,j}(\mbB_{j}^{(t)} \mid \mbB_{-j}^{(t+1,t)}, \boltheta^{(t)})$ by the minorize-maximize principle \citep{lange2016mm}. 
In addition, the maximizer of $\widetilde{\mathcal{M}}_{\lambda,j}(\cdot\mid \mbB_{-j}^{(t+1,t)}, \boltheta^{(t)})$ has the closed form
\begin{equation}\label{eq:B_update}
{\mbB}^{(t+1)}_j  = \max \left(1 - \frac{\lambda \tau_j}{ \|\mbu^{(t)}\|_2}, 0\right) \mbu^{(t)}, 
\end{equation}
where $\mbu^{(t)} = \mbB_{j}^{(t)} + \tau_j \nabla \mathcal{M}_{0,j}(\mbB_{j}^{(t)} \mid \mbB^{(t+1,t)}_{-j}, \boltheta^{(t)}).$
\end{theorem}

The theoretical range for $\tau_j$ is not used in our implementation. Instead, we select $\tau_j$ using an Armijo-type backtracking line search which allows us to consider larger step sizes $\tau_j$ while maintaining the descent property described in Theorem \ref{prop:descent}. 
To obtain the complete $(t+1)$th iterate of $\mbB$, we compute \eqref{eq:B_update} for $j \in [p]$ in a random order. This randomized approach is similar to the random permutation cyclical coordinate descent algorithm of \citet{wright2020analyzing}. After all $p$ rows of $\mbB$ are updated once, we proceed to the next E-step, letting $\boltheta^{(t+1)}$ denote the resulting $\{\delta_r^{(t+1)}, \bolbeta_{r1}^{(t+1)}, \dots, \bolbeta_{rM}^{(t+1)}\}_{r\in[R]}$. It is important to note that by cycling through the rows of $\mbB$ only once, $\mbB^{(t+1)}$ is not, in general, the argument maximizing $\mathcal{M}_\lambda(\cdot \mid \boltheta^{(t)})$. If we repeatedly cycled through all $p$ rows, the iterates would eventually converge to a global maximizer of $\mathcal{M}_\lambda(\cdot \mid \boltheta^{(t)})$. However, we found this approximation scheme to be more computationally efficient than solving the M-step exactly at each iteration, and we can easily verify that it ensures ascent.

\begin{lemma}\label{lemma:ascent}
As long as each $\tau_j$ is chosen according to Theorem \ref{prop:descent} (or by backtracking line search), the objective function from \eqref{eq:penalized_obs_loglik} evaluated at $\boltheta^{(t+1)}$ is guaranteed to be no less than the objective function from \eqref{eq:penalized_obs_loglik} evaluated at $\boltheta^{(t)}$. That is, the sequence of iterates $\{\boltheta^{(t)}\}_{t=1}^\infty$ generated by Algorithm \ref{alg:EM} monotonically increase the value of the objective function from \eqref{eq:penalized_obs_loglik}.
\end{lemma}
Lemma \ref{lemma:ascent} relies on the fact that our algorithm is an instance of the expectation conditional-maximization algorithm \citep{meng1993maximum}.  We summarize the entire algorithm in Algorithm \ref{alg:EM}. To accommodate $\mathcal{H}_\lambda$ rather than $\mathcal{G}_\lambda$, we would need only replace $\|\mbb\|_2$ in \eqref{eq:B_update} with $\sum_{r=1}^R \|\mbb_r\|_2$ where $\mbb = (\mbb_1^\top, \dots, \mbb_R^\top)^\top$ with each $\mbb_j \in \mathbb{R}^{C}$. We would then apply an updating equation like \eqref{eq:B_update} based on the subvectors of $\mbu^{(t)}$ corresponding to each $\mbb_r$.

\begin{algorithm}[t]
    \begin{enumerate}
        \item Initialize $\boltheta^{(0)} = \{\delta_r^{(0)}, \bolbeta^{(0)}_{1r}, \dots, \bolbeta^{(0)}_{Mr}\}_{r \in [R]}$ where $\delta^{(0)}_r > 0$ is the probability that $Z=r$ (i.e., $\sum_{r=1}^R \delta^{(0)}_r = 1)$ and $\bolbeta^{(0)}_{mr} \in \mathbb{R}^{p \times c_m}$  such that for $(r,m) \in [R] \times [M].$
        \item  %When $\lambda = 0$, repeat steps (a), (b), and (c) for $t= 0,1,\ldots$, until convergence;  when $\lambda > 0$ 
        Repeat the following steps for $t= 0,1,\ldots$, until convergence. 
        \begin{enumerate}
            \item E-step:
             For each $r \in [R]$ and $i \in [n]$, compute the conditional probability that $Z_i=r$,
                $$\pi_{ir}^{(t)} = \dfrac{ \delta_r^{(t)} f_r(\mbY_i\mid\mbX_i =\mbx_i,\boltheta_r^{(t)})}{\sum_{s=1}^R\delta_s^{(t)} f_s(\mbY_i\mid\mbX_i =\mbx_i,\boltheta_r^{(t)})},$$
where $f_r(\mbY_i\mid\mbX_i = \mbx_i,\boltheta_r) = \prod_{m=1}^M f_r(Y_{mi}\mid\mbX_i = \mbx_i,\bolbeta_{mr})$ with each $f_r(Y_{mi}\mid\mbX_i = \mbx_i,\bolbeta_{mr})$ being the multinomial probability mass function described in \eqref{multinomial}.
\item[(b)] M-step: Perform (i) and (ii).
\begin{enumerate}
\item[(i)]
For each $j \in [p]$ in a random order, compute
\begin{equation}
\mbB^{(t+1)}_j = \max \left(1 - \frac{\lambda \tau_j}{ \|\mbu^{(t)}\|_2}, 0\right) \mbu^{(t)} 
\end{equation}
where $\mbu^{(t)} = \mbB_{j}^{(t)} + \tau_j \nabla \mathcal{M}_{0,j}(\mbB_{j}^{(t)} \mid \mbB^{(t+1,t)}_{-j}, \boltheta^{(t)})$.
\item[(ii)] For each $r \in [R]$, compute $\delta_r^{(t+1)} = n^{-1}\sum_{i=1}^n \pi_{ir}^{(t)}.$
    \end{enumerate}\end{enumerate}
    \item Output $\widehat\boltheta = \boltheta^{(t+1)}$ at convergence.
    \end{enumerate}
    \caption{Algorithm for maximizing the penalized observed data log-likelihood \eqref{eq:penalized_obs_loglik}}
    \label{alg:EM}
\end{algorithm}
We provide details about our implementation (e.g., initialization scheme and convergence criteria) in Supplementary Material Section S6. Regarding the choice of $R$, we found that cross-validation may not be necessary. In both our simulation studies and real data example, we found that overspecifying $R$ often led to no worse performance than did selecting $R$ by cross-validation. This can be partly explained by the fact that when using our penalties, the penalized EM algorithm automatically forces some $\delta_r$ estimates to be close to zero (e.g., less than $ 10^{-8}$) when $\lambda$ is sufficiently large. Consequently, computing the solution path for our estimator explores both varying levels of sparsity in the regression coefficients and implicitly, various values of $R$. See Section \ref{sec:overspecify_R} for further details. 

\section{Statistical analysis of penalized M-step}\label{sec:theory}

To better understand the performance of our method, we study the statistical error involved in the penalized M-step of Algorithm 1 from Section \ref{subsec:ECM_alg}. In the finite-sample analysis of the maximizer of $\mathcal{M}_\lambda(\mbB\mid  \boltheta^{(t)})$, 
%or the closely related one-step blockwise gradient ascent estimator used in our algorithm, 
it is very challenging to establish uniform concentration inequalities about the stochastic objective function $\mathcal{M}_\lambda(\mbB\mid  \boltheta^{(t)})$ and its gradient $\nabla\mathcal{M}_\lambda(\mbB\mid  \boltheta^{(t)})$, which depend on the estimates $\boltheta^{(t)}$. For example, a theoretical study of the EM algorithm may require sample splitting; given a total of $n$ samples and $T$ iterations, the sample-splitting EM algorithm would use $T$ subsets of size $n/T$ to break the dependence of $\mbB^{(t+1)}=\argmax_{\mbB}\mathcal{M}_\lambda(\mbB\mid  \boltheta^{(t)})$ on $\boltheta^{(t)}$. See \citet{balakrishnan2017statistical} and \citet{ zhang2020estimation} for examples. Because of this challenge, we leave the finite-sample statistical analysis of our penalized EM algorithm as future research. Instead, we study an idealized estimator $\widehat{\mbB}^{\dagger}$ from a modified version of $\mathcal{M}_\lambda(\mbB\mid  \boltheta^{(t)})$ which replaces the $\pi_{ir}^{(t)}$ with the ``oracle'' information $Z_{ir}$. 
That is, by analyzing the maximizer of the penalized conditional log-likelihood of $(\mbY,Z)\mid\mbX$ in \eqref{eq:lik_Z}, we derive bounds that are meant to illustrate how $p, c_1, \dots, c_M$, and the sparsity of the $\bolbeta_{mr}$ affect estimation of the regression parameter $\mbB$ in an idealized scenario. 

Throughout this section, let $|\mathcal{A}|$ denote the cardinality of a set $\mathcal{A}$. Similarly, let $\|\mbA\|_{1,2} = \sum_{k} \|\mbA_{k,:}\|_{2}$ be the norm which sums the Euclidean norms of the rows of its matrix-valued argument. 
To simplify matters, we treat $\boldsymbol{X} = (\mbx_1, \dotsm \mbx_n)^\top \in \mathbb{R}^{n \times p}$ as nonrandom and standardized such that $\sum_{i=1}^n \boldsymbol{X}_{i,j}^2 = n$ for $j \in [p]$.  We focus on the penalized estimator using the global variable selection penalty $\mathcal{G}_\lambda(\mbB)$. In the Supplementary Material Section S6, we discuss how similar results could be obtained under the penalty $\mathcal{H}_{\lambda}.$ To avoid cumbersome tensor notation and operators, we re-define $\mbB=\mbB_{(1)}\in\mbbR^{p\times RC}$ as the matrix parameter in this section. 

The estimator we study, $\widehat\mbB^\dagger$, is defined formally as 
\begin{equation}\label{eq:ideal_estimator}
 \argmin_{\mbB \in \mathbb{R}^{p \times RC}}  \left\{ - \frac{1}{n}\sum_{i=1}^n\sum_{r=1}^R \left[ Z_{ir}\sum_{m=1}^M\log\left\{f_r(Y_{mi}\mid\mbX_i = \mbx_i,\bolbeta_{mr})\right\}\right]  +   \mathcal{G}_\lambda(\mbB) \right\}.
\end{equation}
where $Z_{ir} = \mathbf{1}(Z_i = r)$ for $(i,r) \in [n] \times [R]$. The above estimator hence does not depend on $\boltheta^{(t)}$. We will treat $R$ as a fixed and allow $M$, $c_1, \dots, c_m$, $p$, and $n$ to tend to infinity. Our objective is to establish an error bound on $\widehat\mbB^\dagger - \mbB^\dagger$ where we define $\mbB^\dagger = \argmin_{\mbB \in {\huge \boldsymbol{\pi}^*}} \mathcal{G}_\lambda(\mbB)$ with ${\huge \boldsymbol{\pi}^*}$ denoting the set of all $\mbB$ which lead to the true probabilities $\mbP(\mbx)$ for all $\mbx \in \mathbb{R}^p$. In the Supplementary Material, we show that $\mbB^\dagger$ is uniquely defined, does not depend on $\lambda$, and that for each irrelevant predictor, the corresponding row $\mbB_{j,:}^\dagger \in \mathbb{R}^{RC}$ is zero. Thus, define $\mathcal{S} = \{j: \mbB^\dagger_{j,:} \neq 0, j \in [p]\}$ and $\mathcal{S}^c := \{j: \mbB^\dagger_{j,:} = 0, j \in [p]\}$ as the set of relevant and irrelevant predictors, respectively.

Before we describe the needed conditions and assumptions, a few comments are in order. First, the estimator in \eqref{eq:ideal_estimator} is distinct from the estimator which estimates each $\bolbeta_{mr}$ or even $(\bolbeta_{1r}, \dots, \bolbeta_{Mr})$ separately across $r \in [R]$. The penalty we use, $\mathcal{G}_\lambda$, necessarily ties the estimators together so that even in the case that the $Z_i$ are known, all components are estimated jointly. Second, the $Z_i$'s are random so our statistical analysis must consider the joint distribution of the $Z_i$ and the responses. %Third, this theory is foundational to a more involved theory for the solution to our regularized EM algorithm. Such theory would need to replace the $Z_{ir}$ with their conditional estimates, and would necessarily build upon the proof techniques and arguments developed here. 
Finally, we mention that the estimator in \eqref{eq:ideal_estimator} is also relevant to the emerging literature on regression with heterogeneous sub-populations. In that context, $Z_i$ is the indicator that a subject belongs to a particular sub-population (hence observable), and then our estimator from  \eqref{eq:ideal_estimator} is adjusting for $Z$ in order to harmonize the data \citep{fortin2017harmonization} while our penalty $\mathcal{G}_\lambda$ can identify homogeneous structures across sub-population \citep{tang2016fused}.

Our first assumption is a new version of the restricted eigenvalue condition which applies to data generated from a mixture of regressions model. This assumption will depend on $n_{\rm min} = \min_{r \in [R]} \sum_{i=1}^n Z_{ir}$, the minimum number of samples observed across the $R$ latent states. Let
$\Delta = (\tilde\Delta_{11}, \dots, \tilde\Delta_{M1}, \tilde\Delta_{12}, \dots, \tilde\Delta_{MR}) \in \mathbb{R}^{p \times RC}$
with $\tilde\Delta_{mr} \in \mathbb{R}^{p \times c_m}$ for $(r,m) \in [R] \times [M]$, and let $\{\sum_{i=1}^n \Psi^\dagger_{mr}(\mbx_i) \otimes \mbx_i \mbx_i^\top\}$ be the Hessian of $f_r(Y_{mi} \mid \mbX_i = \mbx_i, \bolbeta_{mr}^\dagger)$ with respect to the vectorization of its argument at $\bolbeta_{mr}^\dagger$. The exact form of the positive semidefinite matrix $\Psi^\dagger_{mr}(\mbx_i) \in \mathbb{R}^{c_m \times c_m}$ is given in the Supplementary Material. 
In addition, let $\mathcal{N}(n_{\rm min}) = \{\mathcal{N}_1, \dots, \mathcal{N}_R\}$ denote the set of $R$-element partitions of $[n]$ where $\min_{r \in [R]}|\mathcal{N}_r| \geq n_{\rm min}$.
\begin{itemize}
\item[] \textbf{A1. (Restricted eigenvalue condition)} There exists a constant $k > 0$ such that
$$0 < k \leq \kappa(\mathcal{S}, n_{\rm min}) = \hspace{-5pt}\inf_{(\Delta, \mathcal{A}) \in \mathbb{C}(\mathcal{S}) \times \mathcal{N}(n_{\rm min})}\sum_{m=1}^M \sum_{r=1}^R  \frac{\tilde\Delta_{mr}^\top\left[\sum_{i\in \mathcal{A}_r} \{\Psi^\dagger_{mr}(\mbx_i) \otimes \mbx_i \mbx_i^\top\} \right]\tilde\Delta_{mr}}{n\|\Delta\|^2_F}, $$
where $\mathbb{C}(\mathcal{S}) = \{\Delta \in \mathbb{R}^{p \times R \sum_{m=1}^M c_m}: \Delta \neq 0, \|\Delta_{\mathcal{S}^c, :}\|_{1,2} \leq 3 \|\Delta_{\mathcal{S}, :}\|_{1,2}\}.$
\end{itemize}
Loosely speaking, one could expect the restricted eigenvalue condition to hold if (i) $n_{\rm min}$ is not too small relative to $n$, (ii) that the probabilities ${\rm Pr}(Y_m = j_m \mid Z = r, \mbX = \mbx)$ for $j_m \in [c_m]$ are bounded away from zero and one for all $(m,r) \in [M] \times [R]$ and all $\mbx \in \mathbb{R}^p$, and (iii) the relevant predictors (indexed by $\mathcal{S}$) are neither too correlated with one another, nor with unimportant predictors.

Explicitly, the quantity $\kappa(\mathcal{S}, n_{\rm min})$ is a function of $n_{\rm min}$ and $\mathcal{S}$. The minimum sample size $n_{\rm min}$ determines the space over which the infimum is taken with respect to $\mathcal{A}$ as it defines $\mathcal{N}(n_{\min})$. If $\Delta$ were fixed, taking the infimum with respect to $\mathcal{A}$ would correspond to finding the worst possible -- in the sense of minimizing $\kappa(\mathcal{S},n_{\rm min}$) -- partition of the subjects into the $R$ latent states with each state having at least $n_{\rm min}$ subjects. The set $\mathcal{S}$, in contrast, defines the set $\mathbb{C}(\mathcal{S}).$  Implicitly, $\kappa(\mathcal{S}, n_{\rm min})$ is a function of $R, M$, and the $\delta_r$'s. In fact, to use this quantity in our proofs, we need the following additional assumption. 
\vspace{-5pt}
\begin{itemize}
\item[] \textbf{A2. ($\delta_r$ bounds)} There exists a constant $v \in (0, 1/2)$ such that $v \leq \delta_r  \leq 1 - v$ for $r \in [R].$\vspace{-5pt}
\end{itemize}
Lastly, we also make an assumption about the data generating process. \vspace{-5pt}
\begin{itemize}
    \item[] \textbf{A3. (Data generating process)} The data $\{(Z_i, \mbY_i)\}_{i=1}^n$ are independent with $\Prob(Z_i=r) = \delta_r,$ and  $f(\mbY_i\mid \mbX_i=\mbx_i, Z_i=r, \boltheta) =  \prod_{m=1}^{M}f_r(Y_{mi}\mid \mbX_i=\mbx_i, \bolbeta_{mr}^\dagger)$
    where $f_r$ is as defined in \eqref{multinomial} for $r \in [R].$
\vspace{-.05in}
\end{itemize}
We are ready to state our main result, which will depend on a sample size condition, \textbf{C1}. \vspace{-10pt}
\begin{theorem}\label{thm:error_bound}
Let $k_1 > 2$ and $k_2 > 2$ be fixed constants. Suppose \textbf{A1}--\textbf{A3} hold. If $\lambda = \{\sum_{m=1}^M c_m /n\}^{1/2} + \{4M k_2 \log p /n\}^{1/2}$, then for $n$ sufficiently large such that condition \textbf{C1} holds,
$$\|\hat\mbB^\dagger - \mbB^\dagger\|_F \leq \frac{3 k_1}{\kappa(\mathcal{S}, n_{\rm min})}\left( \sqrt{\frac{|\mathcal{S}|\sum_{m=1}^M c_m}{4 n}} + \sqrt{ \frac{k_2|\mathcal{S}| M \log p}{n}}\right)$$
with probability at least $1 - p^{1 - k_2/2} - \sum_{r=1}^R {\rm exp}[-2n\{\delta_r - (n_{\min} - 1)/n\}^2]$.
\end{theorem}

% \textcolor{red}{Shall we use $I_n(n_{\min},\delta_r)\leq\exp\{-2n(\delta_r-\dfrac{n_{\min}}{n})\}$ (Hoeffding's inequality)?}

The condition on the sample size (\textbf{C1}) needed for the result of Theorem \ref{thm:error_bound} is as follows. 
\begin{itemize}
\item[] \textbf{C1. (Sample size)} Let $\phi_n = \max_{i \in [n]}\|\mbx_i\|_2  k_1 \lambda \sqrt{54 |\mathcal{S}|} /\{2\kappa(\mathcal{S}, n_{\rm min})\}$ where $\lambda$ is as prescribed in Theorem \ref{thm:error_bound}. The sample size $n$ is sufficiently large with respect to $p, c_1, \dots, c_M, M, k_1, k_2,$ and $|\mathcal{S}|$ such that $e^{-\phi_n} + \phi_n - 1 - \phi_n^2/k_1 > 0.$
\end{itemize}

Our error bound illustrates how our idealized M-step scales with respect to the number of response variables and number of categories per response. It is especially instructive to compare the error bound in Theorem \ref{thm:error_bound} to that from \citet{molstad2020likelihood}. Recall that \citet{molstad2020likelihood} propose an estimator of the regression coefficient tensor under the vectorized model. Unsurprisingly, their error bound scales in $\prod_{m=1}^M c_m$ rather than $\sum_{m=1}^M c_m.$ This can be explained by the fact that the regression coefficient tensor of interest under the vectorized model has $p\prod_{m=1}^M c_m$ elements whereas under our model assumptions, there are only $R\hspace{1pt}p\sum_{m=1}^M c_m$ unknown regression coefficients. %These results may suggest that when $M$ is reasonably large, the method of \citet{molstad2020likelihood} will perform poorly whereas our method may perform well as long as \eqref{LatentMixtureModel} with small $R$ serves as a reasonable approximation to $\mbP$. 

\section{Simulation studies}\label{sec:sim}

\subsection{Data generating models and competing methods}\label{subsec:DataGen}
In this section, we compare our method for estimating conditional probability tensors to numerous alternative approaches. We consider data generating models with $R$, the training sample size, the mixture probabilities, and the magnitude of entries in the $\bolbeta_{mr}$'s varying. For a given $R$, we generate $n$ independent realizations of the predictor $\mbX \sim {\rm N}_{100}(0, \bolSigma)$ where $\bolSigma_{j,k} = 0.5^{|j-k|}$ for $(j,k) \in [p] \times [p]$
and generate $Y_1, \dots, Y_4$ from
$${\rm Pr}(Y_1 = j_1, \dots, Y_4 = j_4\mid \mbX = \mbx) = \sum_{r=1}^{R} \delta_{r} \big\{\prod_{m=1}^4 {\rm Pr}(Y_m = j_m \mid \mbX = \mbx, Z = r)\big\}$$
where 
$ {\rm Pr}(Y_m = j_m \mid \mbX = \mbx, Z = r) =  {\rm exp}(\mbx^\top\bolbeta_{mr{j_m}})/\{\sum_{k=1}^{4}{\rm exp}(\mbx^\top\bolbeta_{mrk})\}$ for $ m \in [4],j_m \in [4],$ and $r \in [R].$
We set $M = 4$ and $c_1 = \cdots = c_4 = 4$ throughout. When $R=2$, we set $\delta_{2} = 1 - \delta_{1}$, whereas when $R =3$, we set $\delta_{2} = 2/3 - \delta_{1}$ and $\delta_{3} = 1/3.$  In both cases, we consider various values of $\delta_1$. In each setting we consider, we randomly select five predictors to affect the conditional probability tensor. Each of the five corresponding rows of the $\bolbeta_{mr}$'s has entries which are drawn independently from ${\rm N}(0, \sigma^2_{\beta}).$ To select tuning parameters, we generate a validation set of size 200 from the same data generating model. To quantify performance of the various estimators, we also generate a testing set of size 1000. 

We consider multiple versions of our method with candidate $R \in [4]$ (denoted \texttt{Mix-1} through \texttt{Mix-4}) and $\mathcal{P}_\lambda$ taken to be $\mathcal{G}_\lambda$. We compare these variations of our method to two methods which implicitly assume independence: fitting separate multinomial logistic regression models to each response with  (a) group-lasso penalty on the rows of each of the regression coefficient matrices (\texttt{Sep-Group}) and (b) the $L_1$-norm penalty applied the regression coefficient matrices (\texttt{Sep-L1}).  \texttt{Sep-Group} is equivalent to our method with $R = 1$ and $\mathcal{G}_\lambda (\mbB)$ replaced with $\sum_{m=1}^M \lambda_m \sum_{j=1}^p \|[\bolbeta_{m1}]_{j,:}\|_{2}$. While it may seem natural to want to compare to the vectorized modeling approach, which treats all possible category combinations as a $4^4$-dimensional multinomial, this is not possible since when even with $n = 450$ (the largest training sample size we consider), the probability that all category combinations are observed is nearly zero in every scenario.  We omit the method of \citet{molstad2020likelihood} from our comparisons because their software cannot be applied with $M > 3$. Moreover, their method is meant to identify which predictors are irrelevant, only affect the response marginal distributions, or affect high-order associations. In comparison, our method is a more general purpose approach for fitting $(\mbY \mid \mbX).$

For all considered methods, tuning parameters are chosen to minimize the negative log-likelihood evaluated on the validation set. For the methods which fit separate models, tuning parameters are chosen for each response separately. We evaluate the performance by calculating the square-root average Kullback-Leibler divergence on the testing set. %$\{n_{\rm test}^{-1} \sum_{i=1}^{n_{\rm test}} \sum_{j_1 = 1}^{c_1} \cdots \sum_{j_M = 1}^{c_M}\hat{P}_{j_1 \dots j_M}(\mbx_i)\log(\hat{P}_{j_1 \dots j_M}(\mbx_i)/P_{j_1 \dots j_M}(\mbx_i))\}^{1/2},$ where $\mbP(\mbx_i)$ is the true probability mass function of $\mbY\mid\mbX = \mbx_i$ and $\hat{\mbP}(\mbx_i)$ is the estimate based on a particular fitted model. 

%In the Supplementary Material, we also report results using the Hellinger distance as a performance metric.

\vspace{-10pt}
\subsection{Results}
In Figure \ref{R2_KL}, we display the square-root average Kullback-Leibler divergences on the testing set for each of the six methods with $R=2$. Focusing first on the top row, where coefficients tend to have smaller magnitude, we see that when $n = 75,$ there is only minor differences between methods. When $n = 150$ or $n = 300$, however, we notice that \texttt{Mix-2}, \texttt{Mix-3}, and \texttt{Mix-4} all tend to outperform the other competitors. This is not surprising given that each of these methods contains the model with the true number of components as a special case. Another interesting aspect of these  results is that as $\delta_1$ approaches 0.5, the differences in performance become more apparent. This too is not surprising because when $\delta_1 = 0.1$, \texttt{Mix-1} could serve as a reasonable approximation to $\mbP$. Results are effectively the same when $\sigma_{\beta} = 2$, except that differences between the methods are more pronounced even in the case that $n = 75$ or $n = 150.$

\begin{figure}[t]
\centering
%\makebox[\linewidth]{(a) $\sigma_{*\beta}^2 = 1, R = 2$}\\
\includegraphics[width=.8\textwidth]{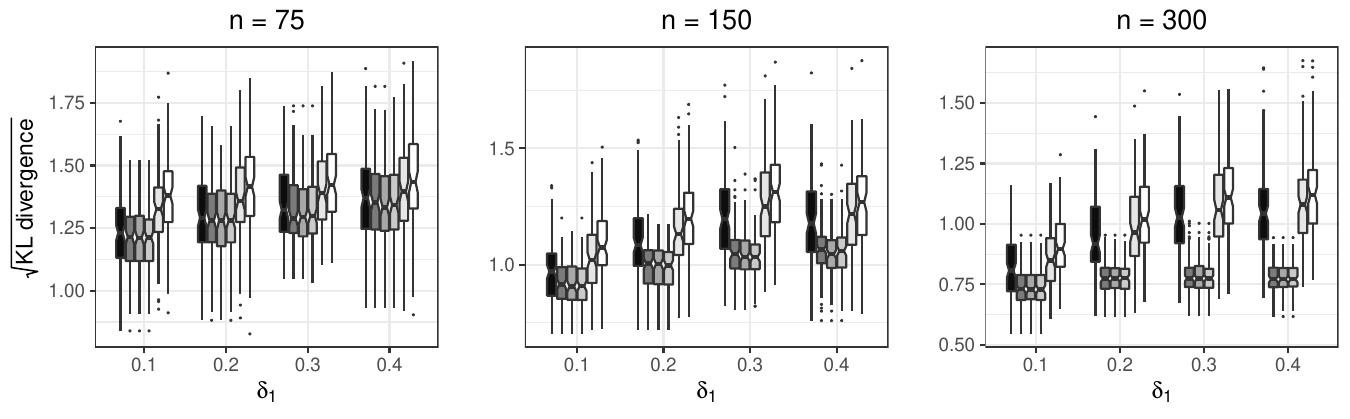}\\
\includegraphics[width=.8\textwidth]{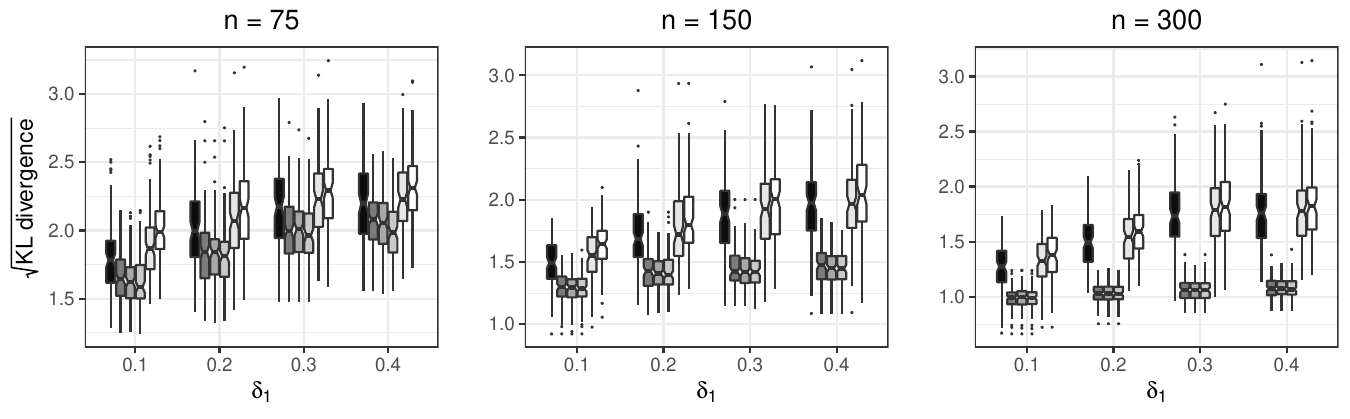}\\
\includegraphics[width=.8\textwidth]{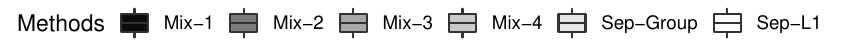}
\caption{The square-root average Kullback-Leibler divergence evaluated on the testing set for the six considered methods. Results displayed are from 100 independent replications with (top row) $(\sigma_{\beta}, R) = (1,2)$ and (bottom row) $(\sigma_{\beta}, R) = (2,2)$.}\label{R2_KL}
\end{figure}

We display analogous results for the data generating model with $R = 5$ in Figure \ref{R5_KL}. Here, $(\delta_1, \dots, \delta_5) = (\delta_1, (1 - \delta_1)/4, \dots (1 - \delta_1)/4).$
Overall, we see results similar to those when $R =2$. Specifically, when $n = 300$ and coefficient magnitudes are (relatively) small, there is little distinction between the methods. As $n = 450$, we see a more clear seperation between the versions of our method with different numbers of mixture components. Notably, we see that as $\delta_1$ increases, \texttt{Mix-2} performs more similarly to \texttt{Mix-5} and \texttt{Mix-7}. This makes sense as when $\delta_1 = 0.67$, four of the five components make relatively small contributions to $\mbP.$ Just as in the case with $R =2$, the magnitude of the regression coefficients $\bolbeta_{mr}$ appears to have an effect as well. Specifically, when $\sigma_\beta = 2$, the difference between estimators is much greater than under similar settings with $\sigma_\beta = 1.$

\begin{figure}[t]
\centering
%\makebox[\linewidth]{(a) $\sigma_{*\beta}^2 = 1, R = 2$}\\
\includegraphics[width=.8\textwidth]{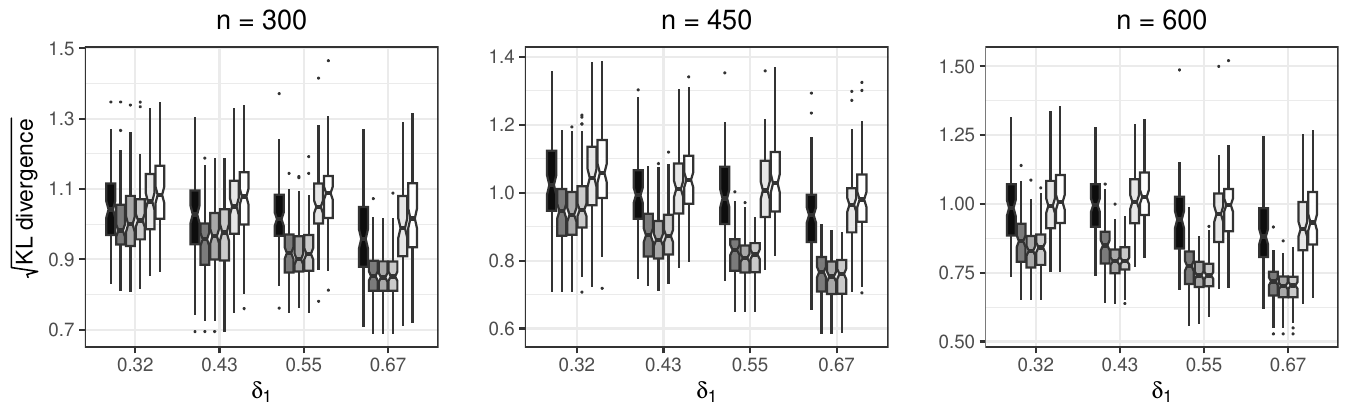}\\
\includegraphics[width=.8\textwidth]{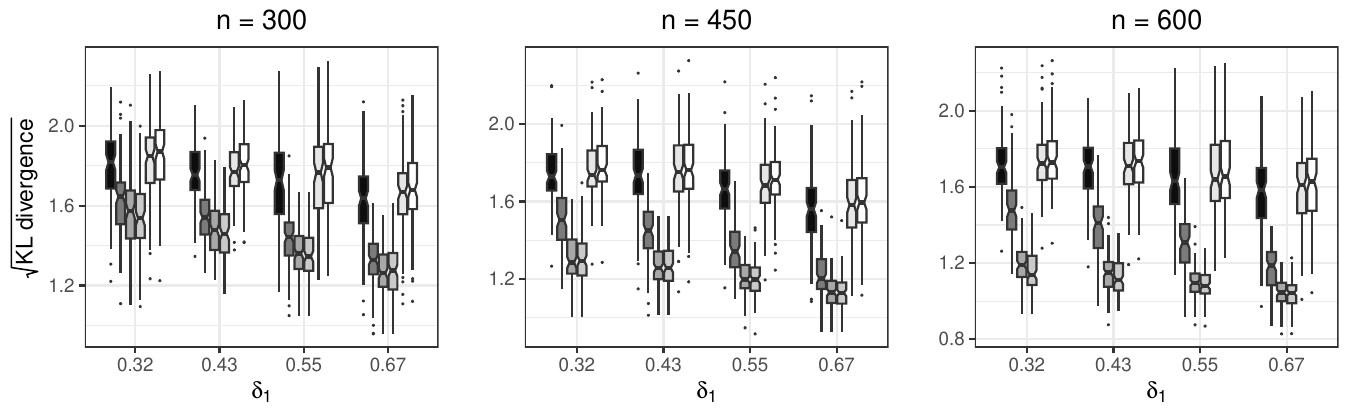}\\
\includegraphics[width=.8\textwidth]{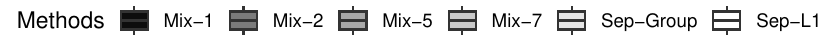}
\caption{The square-root average Kullback-Leibler divergence evaluated on the testing set for the six considered methods. Results displayed are from 100 independent replications with (top row) $(\sigma_{\beta}, R) = (1,5)$ and (bottom row) $(\sigma_{\beta}, R) = (2,5)$.}\label{R5_KL}
\end{figure}

In Section S3 of the Supplementary Material, we present results for the same set of competitors under identical data generating models, but using Hellinger distance as a performance metric. In brief, relative performances are very similar to those based on KL divergence. %However, when $R=3$, the difference between \texttt{Mix-2} and \texttt{Mix-3} is relatively smaller than those observed in Figure \ref{R3_KL}.

% Finally, in Section S2 of the Supplementary Material, we also explore the effect of overspecifying $R$. In brief, we found that even in the case that the true $R = 3$, \texttt{Mix-5}, \texttt{Mix-10}, and \texttt{Mix-15} all perform essentially identically to \texttt{Mix-3}, if not slightly better. 

In the Supplementary Material Section S2 and S3, we include results for numerous additional simulation studies. These include simulation studies with $R \in \{6, 7, 8, 9
\}$, and simulation studies under two types of model misspecification (including when $Z$ depends on $\mbX$). In brief, our method appears reasonably robust against misspecification. 

\subsection{Effect of overspecifying $R$}\label{sec:overspecify_R}
In the simulation results displayed in Figures \ref{R2_KL} and \ref{R5_KL}, we see that overspecifying the number of mixture components seems to have little effect on estimation.  In Figure \ref{R5_KL}, we see that \texttt{Mix-4} tended to perform slightly better than \texttt{Mix-3}, which has the correctly specified number of components. To explore whether more extreme overspecification has a similar effect, we repeated the simulations displayed in the bottom panel of Figure \ref{R5_KL} and added three additional versions of our method with $R \in \{5, 10, 15\}$, which we call \texttt{Mix-5}, \texttt{Mix-10}, and \texttt{Mix-15}, respectively. We display results in Figure S6 of the Supplementary Material. Based on the results, it seems that extreme overspecification of $R$ may even improve estimation accuracy when $n$ is small: \texttt{Mix-15} seemed to perform as well or better than all other methods in each scenario.  However, it is important to note that for the particular tuning parameters chosen for, say, \texttt{Mix-15}, there were often only two nonzero estimates of the $\delta_r$'s. This is one particularly appealing feature of the solution path for fitting \eqref{eq:penalized_obs_loglik}: for large values of $\lambda,$ even when $R$ is large, the solution will have one $\delta_r = 1$ and $\delta_{r'} = 0$ for $r' \neq r.$ As $\lambda$ decreases, eventually a second $\delta_r$ becomes nonzero so that the fitted model effectively has $R=2$. This continues with each additional $\delta_r$ becoming nonzero as $\lambda \to 0.$  We illustrate this phenomenon in Figure \ref{yeast_results} of Section \ref{sec:real}, and discuss this feature of our method in more depth in Supplementary Materials Section S4.

\vspace{-5pt}
\section{Modeling functional classes of genes\label{sec:real}}
In this section, we apply our method to the problem of modeling a gene's functional classes based on both the gene's expression and phylogenetic profile. The dataset we analyzed, which was collected on yeast, was originally studied in \citet{elisseeff2001kernel} and can be downloaded from \url{https://www.uco.es/kdis/mllresources/}. The predictors consist of $p = 103$ components, which are the collection of both the gene's expression and phylogenetic profile. In these data, there are $M = 14$ functional classes (including metabolism, energy, protein synthesis, etc.). Each of the $n = 2417$ genes can be characterized as belonging to multiple functional classes. For example, a gene may affect both metabolism and protein synthesis. Thus, it is natural to treat each functional class assignment as a binary response so that we have $c_1 = \dots = c_{14} = 2.$ 

Because of the relatively large number of response variables, we used the version of our method with the penalty $\mathcal{H}_{\lambda}$ described in \eqref{eq:alt_penalty}. This penalty allows different components of the predictor to be relevant for different values of the latent variable $Z$. In the case that $R=1$, this is equivalent to the penalty $\mathcal{G}_\lambda$, but differs for $R > 1.$

\begin{figure}[t]
\centering
%\makebox[\linewidth]{(a) $\sigma_{*\beta}^2 = 1, R = 2$}\\
\includegraphics[width=0.8\textwidth]{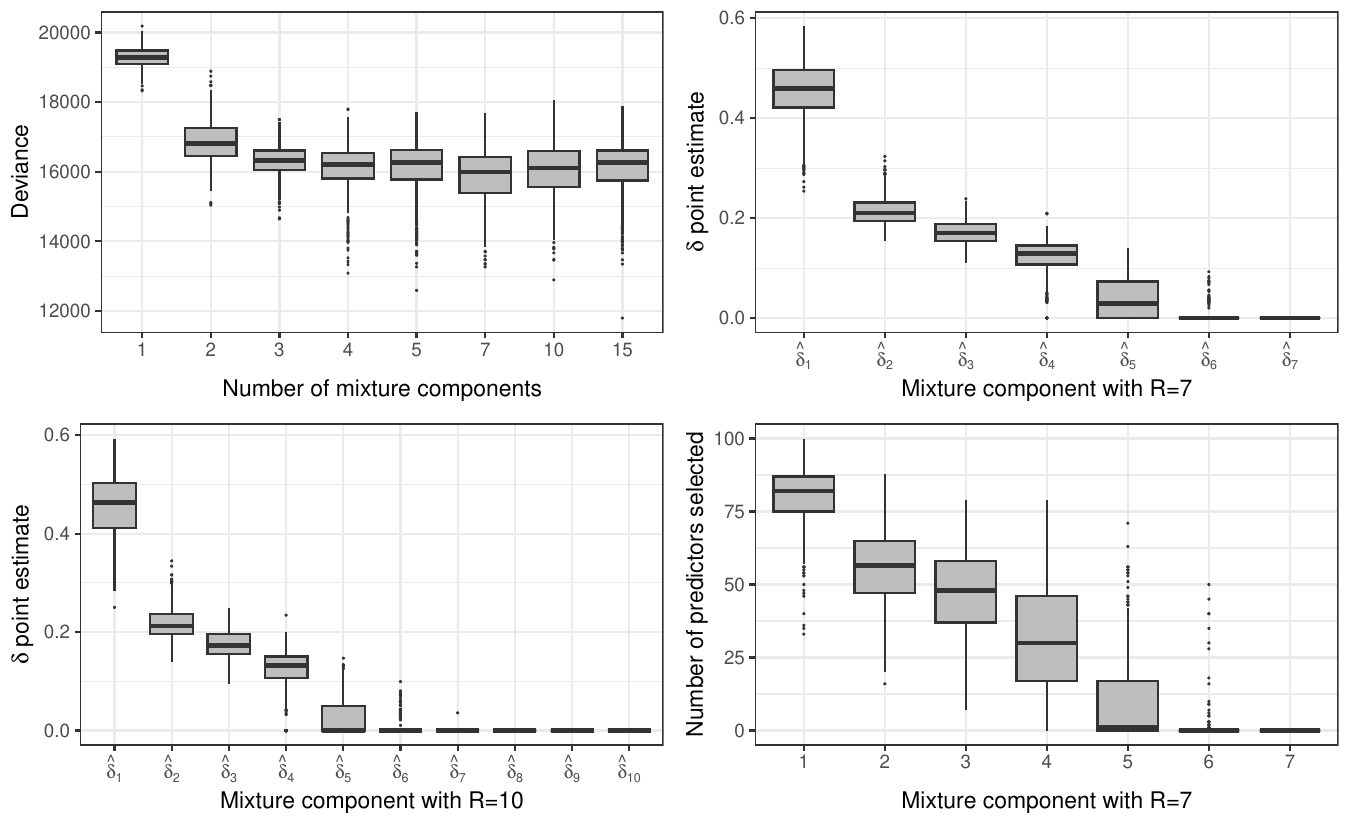}
\caption{Results based on 500 replications: (top left) test set deviance for eight versions of the mixture model with $R \in \{1, 2, 3, 4, 5, 7, 10, 15\},$ (top right, bottom left) ordered estimates of the $\delta_r$ for the version of our method with  $R=7$ and $R=10$, respectively, and (bottom right) the number of predictors selected for each component mixture (in descending order according to the $\hat\delta_r$) with $R = 7$. }\label{yeast_results}
\end{figure}

To compare our method's performance across different choices of $R$, for 500 independent replications we split the data into training, validation, and testing sets of size 1500, 500, and 417, respectively. For $R \in \{1,2,3,4, 5, 7, 10, 15\}$ separately, we fit the model to the training set and selected the 
tuning parameter $\lambda$ by minimizing the negative log-likelihood evaluated on the validation set. Then, we compute the deviance and misclassification accuracy on the testing set for the selected model. In Table 1 of Section S8, we display the average joint error rate and the average testing set deviance for our method with the various number of components. Evidently, $R=7$ performed best in terms of deviance, and was only slight worse (and not significantly different from) $R = 15$ in terms of classification error. %Recall that for each method, we selected the tuning parameters to minimize deviance on the validation set. 
With $R \geq 4$, there is relatively little difference in terms of performance. %This type of finding coheres with the simulation study results which suggest that performance is not hurt when $R$ is over-specified. 

We display the test set deviances in the top left panel of Figure \ref{yeast_results}.  Note that the version of our method assuming independent responses ($R=1$) performs much worse than the versions with $R > 1$; adding even the second mixture component $(R=2)$ decreased testing set deviance by nearly 15\%. In the two rightmost panels, we display the estimated $\delta_r$ and the number of predictors selected in the corresponding $\bolbeta_{mr}$'s with $R = 7$. We see that in general, when $R=7$, the selected model effectively has 5 or fewer $\hat\delta_r$ nonzero. We also notice that the mixture component with the highest probability often had $75$ or more predictors included in the model, whereas those mixture components with smaller probabilities tended to have fewer.  Examining the results with $R = 10$ in the bottom leftmost panel, we again see often only five of fewer $\hat\delta_r$ are estimated to be nonzero.

\vspace{-5pt}
\section{Discussion}\label{sec:discussion}

In this article, we propose a general modeling strategy for the regression of a multivariate categorical response on a high-dimensional predictor based on the population-level tensor rank decomposition of the conditional probability tensor function. %Our approach exploits the connection between the probability tensor rank decomposition and the independence of the response variables conditioning on the predictor and a latent discrete variable $Z$. %This leads to a latent variable interpretation and a mixture generalized linear regression model. Based on an efficient generalized EM algorithm, we are able to obtain a computationally guaranteed solution under group-structured penalization for either global or local variable selection. 
Numerically, our method is shown to perform well with a large number of response variables and large number of categories per response---a setting where many existing competitors fail or cannot be applied. Our results also suggest that our method is insensitive to over-specification of the number of mixtures $R$. %When given a large (potentially over-specified) value $R$, the algorithm can automatically determine the size of some mixture probabilities to be extremely small. 

Based on our theoretical analysis of an idealized and simplified estimator, we conjecture that the convergence rate of the penalized EM algorithm, under suitable assumptions \citep[see, for example,][]{balakrishnan2017statistical,cai2019chime}, is $\sqrt{|\mathcal{S}|M\log p/n}$ when treating $R$ and $c_m$'s as constants. We leave this as a future theoretical study. 
Promising future research directions also include (i) initialization of the algorithm and (ii) alternative convex penalties on the tensor $\mbB$. Regarding (i), methods such as \citet{sedghi2016provable} are applicable to our setting and could potentially improve over the random initialization that we are currently using. For (ii), recent advances on tensor nuclear norm penalties \citep[e.g.,][]{raskutti2019convex} neatly fit in our optimization framework and can take advantage of the tensor construction of $\mbB$.

\bibliographystyle{apalike}
\bibliography{ref_LatentVariable}

\end{document}